# *In Situ* Nanometer-Resolution Strain and Orientation Mapping for Gas-Solid Reactions via Precession-Assisted Four-dimensional Scanning Transmission Electron Microscopy

Yongwen Sun[1†], Ying Han[1†], Dan Zhou[2,3*], Athanassios S. Galanis[4], Alejandro Gomez-Perez[4], Ke Wang[5], Stavros Nicolopoulos[4], Hugo Perez Garza[3], Yang Yang[1,5*]

[1] Department of Engineering Science and Mechanics, The Pennsylvania State University, University Park, PA, 16802, United States.
[2] Leibniz-Institut für Kristallzüchtung (IKZ), Max-Born-Str. 2, 12489 Berlin, Germany.
[3] DENSsolutions B.V., Delft, South Holland, 2628 ZD, Netherlands.
[4] NanoMEGAS SPRL, Rue Émile Claus 49 bte 9, 1050 Brussels, Belgium.
[5] Materials Research Institute, The Pennsylvania State University, University Park, PA, 16802, United States.

[†] These authors contributed equally to this work.
[*] Email of corresponding authors:
  dan.zhou@ikz-berlin.de (D.Z.), yangyang@alum.mit.edu (Y.Y.)

**Abstract**

Chemomechanical interactions in gas or liquid environments are crucial for the functionality and longevity of various materials used in sustainable energy technologies, such as rechargeable batteries, water-splitting catalysts, and next-generation nuclear reactors. A comprehensive understanding of nanoscale strain evolution involved in these processes can advance our knowledge of underlying mechanisms and facilitate material design improvements. However, traditional microscopy workflows face challenges due to trade-offs between field of view (FOV), spatial resolution, temporal resolution, and electron beam damage, particularly in gas or liquid environments. Here, we demonstrate *in situ* nanometer-resolution strain and orientation mapping in a temperature-controlled gas environment with a large FOV. This is achieved by integrating a microelectromechanical system (MEMS)-based closed-cell TEM holder, precession-assisted four-dimensional scanning transmission electron microscopy (4D-STEM), and a direct electron detector (DED). Using the strain evolution during zirconium initial oxidation as a case study, we first outline critical strategies for focused ion beam (FIB) gas-cell sample preparation and gas-phase TEM workflows to enhance experimental success. We then show that integrating DED with precession electron diffraction (PED) and optimizing gas pressure substantially improve the quantity and quality of the detected Bragg peaks in nano-beam electron diffraction (NBED) patterns, enabling more precise strain measurements. Furthermore, we introduce a practical protocol to pause the reactions, allowing sufficient time for 4D-STEM data collection while ensuring the temporal resolution needed to resolve material dynamics. Our methodology and workflow provide a robust framework for quantitative analysis of chemomechanical evolutions in materials exposed to gas or liquid environments, paving the way for improved material design in energy-related applications.

# 1. Introduction

Chemomechanical interactions are prevalent in materials used in advanced energy systems[1,2]. In some cases, these interactions can have detrimental effects. For instance, in rechargeable lithium-ion batteries, the anisotropic volumetric changes during charging and discharging induce aggressive internal strain that accelerates electrode degradation and capacity fade[1,3–6]. Similarly, zirconium alloy cladding, which encases fuel in commercial nuclear reactors, suffers from oxidation during reactor operations. The stress accumulated in the oxide can lead to crack formation, further accelerating oxidation[7]. Conversely, the interplay between chemistry and mechanical behavior has also been exploited to optimize material performance or enable novel functionalities, such as in strain-engineered catalysts[8,9] and mechanical energy harvesters[10].

An in-depth understanding of the nano-structural evolution during chemomechanical interactions is critical for enhancing material performance and durability, especially under harsh conditions like high temperatures and corrosive gas or liquid environments. However, a major challenge has been the lack of techniques capable of directly observing strain evolution at nanometer resolution in such environments.

To address this gap, substantial research has been dedicated to probing strain distribution using techniques such as X-ray diffraction[11–13], neutron diffraction (ND)[14,15], electron backscatter diffraction (EBSD)[16,17], Raman spectroscopy[18–20], and TEM or STEM[21–23]. High-energy synchrotron X-ray is valuable for acquiring local strain within bulk materials and is further able to characterize full lattice strain tensor *in situ* and in 3D at a resolution of approximately 22 nm[11,13]. ND is well suited for operando strain measurements, allowing for location-specific characterization[15], but the spatial resolution (typically 1 mm) is lower than characterization techniques using X-ray or charged particles. EBSD can identify regions of concentrated strain from dislocation accumulation at sub-micron spatial resolution; however, it is limited to surface characterization and struggles with quantifying the strain magnitude[16]. Raman spectroscopy provides surface strain information by analyzing peak shifts and broadening due to inelastic photon scattering, with recent developments achieving submicron spatial resolution in operando 3D Raman spectroscopy[24].

Among these methods, TEM/STEM stands out for offering nanometer or sub-nanometer spatial resolutions[21] and through-thickness analysis. Strain characterization in TEM/STEM typically falls into two categories: real-space based methods and reciprocal-space based methods. The former leverages atomic column displacements observed in atomic-resolution images, quantified through techniques like geometric phase analysis (GPA)[25–28] or tracking of individual atomic column positions[29–31]. While these methods enable sub-nanometer spatial resolution analysis, they face several limitations. First, they normally have a limited FOV (typically around 100 nm) and require large beam currents, which increase electron beam damage. Second, they are sensitive to sample drift (especially in HR-STEM due to longer frame times compared to HR-TEM), sample bending, and variations in sample thickness. Third, for HR-TEM, the lens imperfections can cause discrepancies between observed atomic lattice contrast and its actual location[32], while HR-STEM might introduce noise or drift error arising from the scanning process[33,34]. Furthermore, they necessitate precise zone-axis alignment. Local deviations in crystal orientation due to sample bending can impede successful strain mapping. Consequently, real-space imaging-based methods often have limited precision[35] in strain measurement and are primarily useful for detecting significant lattice deformations[36].

Reciprocal-space based methods, particularly, scanning beam electron diffraction techniques in STEM, have been extensively used for strain mapping, especially in semiconductor research. These include convergent beam electron diffraction[37,38] and nano-beam electron diffraction (NBED)[36,39,40]. The advancement of fast cameras such as DEDs has further advanced these techniques, ushering in the era of 4DSTEM[41,42]. In a typical 4D-STEM experiment for strain mapping, a nano-sized electron beam (~ 1 to 10 nm) scans across the sample, recording a nanobeam electron diffraction (NBED) pattern at each point in real space[43]. By analyzing spacing of Bragg reflection disks within these NBED patterns, lattice spacings and the corresponding elastic strain can be calculated[21,35].

Although NBED-based strain mapping has lower spatial resolution than real-space based methods, it offers higher precision of strain. This is because strain-induced shifts in diffraction spots are more pronounced than shifts in atomic column positions in real-space images. Moreover, NBED is less sensitive to sample thickness, aberration variation, or physical sample drift[44], making it more appropriate for large-area measurements. It provides a large FOV of up to 10 μm with nanometer spatial resolution[44,45], while requiring significantly lower electron beam currents[43] than HR-TEM/STEM.

Reliable strain mapping with 4D-STEM hinges on accurate Bragg peak detection[46], which presents several challenges. First, the technique requires precise alignment of the sample to a low-index zone axis, which can be difficult in *in situ* setups with atmospheric pressure gas or liquid environment where single-tilt TEM holders are commonly used. Second, changes in crystal orientation due to defects, bending, or grain rotation can reduce the number of detectable Bragg disks, limiting strain analysis. Third, thicker samples introduce complications such as Kikuchi lines[47], increased dynamical scattering effects and inelastic scattering[48], which can lead to non-uniform and sometimes undetectable Bragg disk fringes[46]. These factors have hindered the use of 4D-STEM for *in situ* strain evolution studies.

The recent development of gas and liquid environmental capabilities in *in situ* TEM[49–52] has opened new research opportunities but also introduced additional challenges for 4D-STEM-based strain and orientation mapping. TEM typically operates under high vacuum to protect the electron source and maintain beam coherence, which is essential for achieving high spatial and energy resolution. Alongside the invention of the electron microscope in the 1930s in Berlin, *in situ* environmental electron microscopy with differential pumping and closed-cell concepts emerged to enable observation of reactions of samples in a non-vacuum environment in a TEM[53]. For a long time, the differential pumping-based approach has been more widely applied due to its easier sample and *in situ* system preparation, despite that its upper limit of pressure is about 20 mbar. The closed cell approach was less used due to the low success rate in preparing a leak-tight cell and a lack of knowledge in proper usage of the closed cell. However, in the past 5 to 10 years, the maturation of industrial-scale production of nano cells based on microelectromechanical systems (MEMS) has significantly improved the simplicity and robustness of close-cell MEMS operation. Utilizing electron-transparent $Si_3N_4$ windows, these nano cells can encapsulate the specimen, gas and circuits for heating and electrical biasing and separate them from the TEM vacuum[54–58]. Moreover, they can reach much higher gas pressures, up to 2 bar for commercially available systems[59], making *in situ* studies more relevant to real-world situations.

Despite these advances, MEMS closed-cell systems introduce additional complexities for strain and orientation mapping. The effective sample thickness increases due to the gas/liquid medium and $Si_3N_4$ windows, degrading the accuracy and robustness of 4D-STEM strain analysis. Additionally, 4D-STEM experiments are inherently time-consuming, as data acquisition is often constrained by camera speed or low electron beam currents. Even with fast DED cameras, collecting a full 4D-STEM dataset can take several minutes[60], whereas chemomechanical processes can lead to significant changes in materials within seconds. This creates a critical need to pause material evolution during experiments to enable high-quality 4D-STEM data acquisition. Furthermore, the single-tilt limitations imposed by gas/liquid tubing requires additional dedicated steps during focused ion beam (FIB) sample preparation to extract and properly align the samples.

In this work, we address these challenges by integrating several recent innovations in electron microscopy, including precession-assisted 4D-STEM, DED cameras, and MEMS-based in situ gas-phase TEM systems.

Precession electron diffraction (PED) is a specialized technique for collecting diffraction patterns, which modifies conventional illumination by dynamically precessing the electron beam around the optical axis. The hollow-cone illumination in PED leads to an integration of diffraction intensities through a beam tilt averaging process, resulting in quasi-kinematical diffraction patterns[61]. PED reduces heterogeneity within the diffraction disks, enhances the capture of higher-order reflections, and facilitates easier zone axis alignment[46,62]. Precession-assisted 4D-STEM can enhance the robustness of strain and orientation mapping, as the precession of the electron beam reduces the sensitivity of diffraction patterns to issues caused by the sample thickness and slight zone axis misalignments[35,62,63]. In the past, 4D-STEM was limited by the slow acquisition speeds of conventional charge-coupled device (CCD) cameras. The recent development of DED facilitates high-speed and high-dynamic-range recording of electron microscopy images[64]. The integration of 4D-STEM with DED can effectively enable much faster data capture at specific electron dose rates while ensuring a suitable signal-to-noise ratio (SNR), thereby boosting the temporal resolution and image quality at the same time[65,66]. Additionally, recent development of MEMS-based *in situ* gas phase TEM holder offers rapid thermal and pressure responses achievable within 5-8 seconds, facilitating the meticulous design of gas and temperature profiles that can either promote or pause the reaction[55]. Furthermore, considerable effort is also dedicated to gas pressure modulation, sample preparation and data acquisition to ensure sample stability and data quality across a large FOV.

By leveraging these advancements, in this work, we develop a systematic and reliable workflow for *in situ* nanometer-resolution strain and orientation mapping in gaseous environments using precession-assisted 4DSTEM. By using initial oxidation of pure Zr as a study case, we outline a comprehensive workflow that covers sample preparation, precession-assisted 4D-STEM data collection strategies, and optimization of the MEMS-based gas and temperature profiles.

## 2 Materials and Methods

*2.1 Materials and sample preparation*

A bulk single-crystal $\alpha$-Zr specimen with [0001] as the surface normal was selected for TEM sample preparation. A Thermo Fisher Scientific (TFS) Helios NanoLab 660 dual-beam FIB was used to extract and prepare an electron-transparent Zr lamella with [0001] as the surface normal, followed by transfer to a DENSsolutions MEMS-based gas-cell chip.

For FIB lamella preparation, a 1 μm platinum protective layer was deposited over the region of interest before milling. A final cleaning step was performed on both surfaces of the lamella using a 2 keV $Ga^+$ beam. During the FIB transfer, a 5 keV electron beam was used for imaging while the $Ga^+$ beam was used exclusively for cutting and depositing, minimizing ion beam damage. Consistent sample tilt angle compensation was applied to facilitate parallel alignment between the lamella and the MEMS-chip surface. Finally, carbon deposition was performed at both lamella edges using a 5 keV $Ga^+$ beam at 0.32 nA to secure the lamella onto the MEMS-chip. Additional detailed procedure of FIB can be found in section 3.1.

*2.2 Electron microscopy characterization*

An FEI-Tecnai-G2 S-TWIN TEM, equipped with a LaB$_6$ electron gun and a CheeTah M3 hybrid pixel direct electron detector (Amsterdam Scientific Instruments), was used for 4D-STEM data acquisition. The CheeTah M3 detector features a Medipix-3 chip bonded to a 200 μm thick silicon sensor, enabling high-speed, high-dynamic-range, frame-based detection. The detector was mounted on a retractable wide-angle port of the microscope. A NanoMEGAS DigiStar P2010 PED system was integrated with the TEM to synchronize beam precession and scanning. And NanoMEGAS Topspin software controlled the hardware, adjusted scanning parameters, and ensured full synchronization of beam precession and scanning with detector recording.

The microscope was operated at 200 kV with a spot size of 9 and a condenser aperture of 30 μm. During precession-assisted 4D-STEM experiments, an electron beam with a small convergence semi-angle of 0.8 mrad scanned the sample, capturing an NBED pattern at each real space position in the selected region. Diffraction patterns were captured at 256 × 256 pixels per frame with a 16-bit dynamical range, optimizing the SNR. The SPED was performed with 100 Hz precession frequency, and a precession angle of 1°. Both PED and DED were instrumental in ensuring accurate strain measurements, as demonstrated in previous research[35,44,65]. The employed DED has a data readout speed of 2000 frames per second. The exposure time used in **Fig. 2**, **Fig. 4,** and **Fig. 6** are 30, 40 and 40 ms, respectively. Scanning step sizes of 12 nm, 25 nm, and 12 nm were used to encompass varying fields of view as discussed in the following **Fig. 2**, **Fig. 4,** and **Fig. 6**. Prior to dataset acquisition, each sample was tilted so that the hexagonal close packed (HCP) Zr crystal was aligned to [0001] zone axis as close as possible.

*2.3* In situ *gas phase TEM set up*

The DENSsolutions Climate Gas Supply System (GSS) and *in situ* gas and heating TEM holder [67–69] were used for *in situ* TEM experiments under an ambient-pressure gas environment. High purity O$_2$ and Ar gases (99.999%) were used as received from Air Liquide. The Climate system enables rapid switching of gas composition, pressure and temperature conditions around the TEM sample, typically within a few seconds, as detailed in a previous report[70]. Gas flows of 0.036 ml$_n$/min of oxygen and 1.964 ml$_n$/min of Ar (resulting in a 1.8% O$_2$/Ar mixture) were first combined at the mixing valve, and then partially directed to the TEM holder, where the inlet and outlet pressure controllers adjusted the flow rate to achieve the required nano-reactor gas pressure ($P_{NR}$) and flow rate ($F_{NR}$)[70], while the remaining gas was vented to the exhaust.

*2.4 Strain and orientation analysis*

In this work, NanoMEGAS Topspin software[71] was used for the strain mapping analysis of the 4D-STEM datasets. The 16-bit raw NBED data was used to calculate the strain. Strain measurement is accomplished by comparing the positions of diffraction spots of the series patterns with the corresponding position of the reference pattern that has been previously acquired from a non-strained area, prior to any *in situ* treatment of the sample. The strain results provide a two-dimensional map of the infinitesimal strain tensor in the plane perpendicular to the beam[27,72,73].

$$\varepsilon = \begin{bmatrix} \varepsilon_{xx} & \varepsilon_{xy} \\ \varepsilon_{yx} & \varepsilon_{yy} \end{bmatrix} \qquad (1)$$

The $\varepsilon_{xx}$ and $\varepsilon_{yy}$ were calculated with respect to the defined coordinate system shown in **Fig. 2a** and **4a**. Specifically, the *x*-axis was chosen to align with the $[1\,0\,\bar{1}\,0]$ of the Zr crystal, while the *y*-axis aligns with the [$\bar{1}2\bar{1}0$] direction. The shear strain components $\varepsilon_{xy} = \varepsilon_{yx}$ correspond to half of the engineering shear strain and follow the sign convention of the engineering stress.

The in-plane hydrostatic strain ($\varepsilon_{hyd}$) was further calculated using the $\varepsilon_{xx}$ and $\varepsilon_{yy}$ maps with the equation:

$$\varepsilon_{hyd} = \frac{1}{2}(\varepsilon_{xx} + \varepsilon_{yy}) \quad (2)$$

Specifically, the $\varepsilon_{xx}$ and $\varepsilon_{yy}$ maps were exported as tiff files from Topspin software and then imported into ImageJ software[74]. The in-plane hydrostatic strain maps were obtained by averaging these two strain components.

The zero-strain lattice spacing reference was obtained by averaging lattice parameters across the entire scanned region for results shown in **Fig. 2** and **Fig. 4**. For results in **Fig. 6**, all the analysis used the averaged lattice parameters across the scan taken before the reaction as the zero-strain reference.

The NBED patterns in the 4D datasets were down-sampled to a 144×144 resolution and converted to 8-bit to facilitate orientation mapping using ASTAR software[75] via template matching. The pixel colors in the orientation maps represent different crystal orientations, following the inverse pole figure color code, which encompasses all possible orientations of the crystal structure. The hexagonal phase of the Zr crystal structure (space group P63/mmc, $a$ = 3.232, $c$ = 5.147, $\gamma$ = 120°) was used to generate the simulated diffraction pattern templates. These templates served as references for indexing the experimental series and generating the orientation maps. The inverse pole figures of the orientation maps are specially cropped to enhance the contrast of the small variation of orientation in the samples (see Supplementary file).

*2.5 Multislice simulation*

The multislice simulation is performed using the abTEM simulation package[76]. To evaluate the effect of crystal out-of-plane tilt on the strain measurement accuracy, we compare the simulated selected area electron diffraction (SAED) patterns of a strain-free pure $\alpha$-Zr with different mistilts (0° to 10°) on the [0001] zone axis. In our simulations, a supercell with dimensions of 177.76 Å ($x$), 179.14 Å ($y$), and 102.94 Å ($z$) was constructed. The electrostatic potential was calculated using an accelerating voltage of 200 kV, and a reciprocal space sampling resolution of 0.02 Å$^{-1}$. The potential was sliced along the beam propagation direction with a slice thickness of 0.25 Å, and the infinite potential projection method was used. Gaussian smoothing with a width of 2 pixels was applied to all the SAED patterns.

## 3. Results and Discussions

*3.1 FIB-based sample preparation for precession-assisted 4D-STEM experiments in gas cell environments*

FIB enables precise sample shaping and positioning on MEMS chips, making FIB lift-out techniques the preferred method for preparing TEM samples from bulk materials for MEMS-based *in situ* experiments[77–80]. Reliable *in situ* TEM sample preparation must preserve the microstructure, and chemistry of the materials, achieving electron transparency while minimizing contamination and intrinsic property degradation.

In this section, we presents how we addressed several key challenges associated with sample preparation for gas cell environments in *in situ* TEM experiments: (1) aligning the sample close to a low-index zone axis; (2) avoiding sample bending and ensuring mechanical support during thinning; (3) preventing damage during assembly; (4) minimizing FIB damage and contamination; (5) mitigating charging effects; and (6) avoiding redeposition and ensuring accurate overlap with the MEMS window.

Strain mapping via 4D-STEM requires precise alignment of the sample near a low-index zone axis—a complementary justification for this requirement is provided in Section 3.2. This alignment is particularly challenging when using a single-tilt TEM holder and working with materials that have small grain sizes. To overcome these, we chose a single crystal α-Zr bulk sample from which to extract specimens. Additionally, we employed an in-plane transfer method that enables us to extract a thin TEM foil whose plane normal is aligned with the surface normal of the bulk sample. The step-by-step procedure for our sample preparation methodology is depicted in **Fig. 1**.

First, using FIB, a small block measuring 30 µm in length, 7-10 µm in width (parallel with the bulk surface), and nearly 2 µm in thickness (perpendicular to the bulk surface) was lifted out from a α-Zr bulk sample whose surface normal is along [0001] direction. Then, the block was transferred and attached to a TEM half grid laid horizontally on the SEM stub (**Fig. 1a**). The half grid was then taken out of the FIB and flipped 90°, thereby the grid was mounted vertically on the SEM stub (**Fig. 1b**) for further thinning in the FIB. This process ensures that the [0001] direction of the thinned lamella is aligned parallel with the surface normal of the lamella. Note that a protective platinum layer of approximately 1 µm thick was deposited over the top of block right before thinning to reduce the "curtain effect"[81] during ion-beam thinning. To avoid bending of the sample during the thinning process, only a small window region on the lamella was thinned, thus the lamella has a U-shape frame which is much thicker than the thinned region, providing the necessary mechanical support to preserve the shape (see the inset of **Fig. 1b**). However, it is worthwhile to note that the distance between the top and bottom $Si_3N_4$ window varies depending on the chip designs, and this gap distance indicates the maximum sample thickness allowed. To ensure the lamella does not damage the MEMS chip during cell assembly, we kept the maximum sample thickness below 2 µm in our experiments. The thinned window in **Fig. 1b** acts as the region of interest (ROI) for *in situ* S/TEM experiments.

Throughout the FIB lamella preparation, $Ga^+$ induced irradiation is inevitable. The energy of the $Ga^+$ beam was progressively reduced step by step from 30 kV to 16, 8, and finally 5 kV during the thinning process. After achieving electron transparency, a final cleaning of 2 kV was applied to minimize ion beam damage to the sample surface.

Upon completing the lamella thinning, the sample was removed from the FIB chamber to flip again thus the half grid was laid horizontally, next to a MEMS chip (**Fig. 1c**).

Avoiding the charging effect during e-beam or ion-beam imaging is critical for the success of the sample preparation and the following *in situ* TEM experiments. Several procedures have been developed to meet this need. A 3 nm carbon coating was sputter deposited to the backside of the chip before the FIB experiments, as shown in **Fig. 1d.** The chip after carbon coating shows a slight color change in the optical microscopy image. Additionally, both the TEM half grid and MEMS chip were secured by conductive 3M™ XYZ Axis Tape on the stub. Also, an extra conductive tape was used to connect the stub and the electrodes on the MEMS chip to help ground the chip. Noted that the tape should not cover the window of the MEMS chip.

Next, the grid and the MEMS chip were loaded in the FIB for transferring the TEM sample to the MEMS chip. During the process of transferring the lamella to the MEMS chip, the easy lift needle was introduced and its tip was gently welded to the lamella using carbon deposition (5 kV 0.32 nA, ion beam). Note that the junction of the tip and the lamella is located at the U-shape section of the lamella (**Fig. 1c**), because the thicker U-shape section served as a protective barrier for the MEMS chip when removing the junction between the needle and the sample by Ga$^+$ beam after the transfer. Note that it is advisable to avoid imaging the thin part with the Ga$^+$ beam during the whole process to prevent Ga$^+$ damage and contamination. In addition, it is critical to ensure the thin region of the lamella overlap with the Si$_3$N$_4$ window of the MEMS chip, and the lamella should be parallel with the chip surface. To secure the lamella to the MEMS chip and prevent potential rolling during later experiments, carbon deposition was applied at least at both sides of the lamella, as demonstrated in **Fig. 1c** and **e**. **Figs. 1f-g** show an example of a rolled-up sample after heating experiment when only one side of the lamella is welded on the chip. The final step involved removing the junction between the needle and the lamella, with special attention to avoid milling a hole in the e-chip that could cause gas leakage.

Even with much carefulness throughout both the preparation and transfer stages, there may still be some undesired sample alignment causing the [0001] zone axis imperfectly aligned parallel to the surface normal of the chip. Typical reasons may include the effect of van der Waals force between the lamella and the chip causing tilting of the lamella, or the misalignment of the chip/grid/FIB-stage with respect to the horizontal plane. Such misalignment can be partially corrected by the single tilt function of the TEM holder before the 4D-SPED experiments. In the next section, we will show that PED could compensate for the minor mistilt of sample for strain mapping purposes.

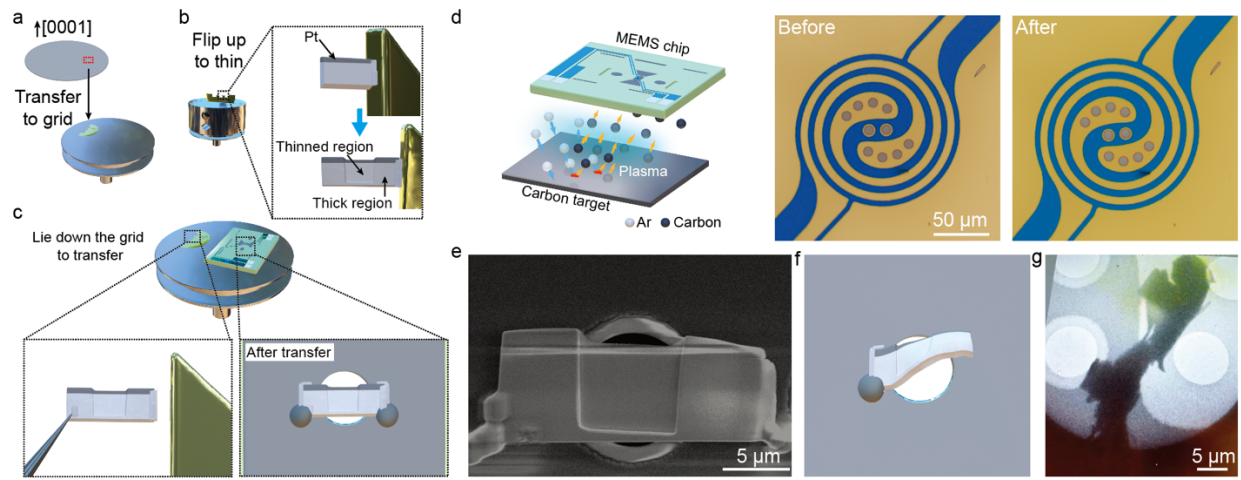

**Fig. 1.** Illustration of the sample preparation procedure. (a) In-plane FIB lift-out of Zr single crystal and transfer to a copper grid laid horizontally. (b) The half grid was flipped 90°, followed by the deposition of a protection Pt cap and Ga$^+$ thinning. The surface normal of the lamella is along the [0001] direction of Zr, and the electron transparent region is surrounded by a thicker U-shape Zr for mechanical support. (c) Using FIB to transfer the foil from the copper grid to the MEMS chip. (d) Schematic drawing showing carbon deposition on the back of the MEMS e-chip to reduce charging effect and optic microscopy images showing the change of the chip before and after carbon deposition. (e) SEM image showing a Zr lamella on the MEMS chip after the transfer process. The round hole beneath the sample is the Si$_3$N$_4$ window. (f) Schematic drawing of a sample that rolls up due to lamella being welded at only one corner. (g) A TEM image showing a rolled-up sample because of welding at just one corner.

*3.2 Precession-assisted 4D-STEM on a gas cell with internal pressure under vacuum*

Strain measurement from a diffraction pattern is based on the inverse relationship between crystallographic plane spacing and the distance between diffraction spots. The accuracy of strain measurements based on the individual Bragg peak detection method relies on two key conditions: (1) the precise determination of the Bragg disk centers/edges, which could be complicated by non-uniform and low SNR intensity distribution in one Bragg disk; and (2) the visibility of a sufficient number of Bragg reflections along two non-parallel directions, which is typically ensured when the sample is aligned to a low-index zone axis. However, achieving these conditions can be challenging, especially in *in situ* gas-cell experiments.

The challenges for the first condition arise from several factors: (a) dynamical scattering or the formation of Kikuchi lines due to interaction between the electron beam and a thick sample[35], (b) a diffuse background caused by the inelastic scattering of electrons with the sample, gas, and the $Si_3N_4$ window[82], and (c) insufficient SNR due to low exposure time or low electron detection efficiency in the detector.

For the second condition, deviations from the desired condition can occur due to (a) dynamical scattering, and (b) the sample mistilt due to local strain, bending, or limitations of sample tilting in a TEM holder.

To improve the accuracy and robustness of strain mapping, we integrated precession-assisted 4D-STEM, DED, and optimized convergence angle to mitigate these challenges.

PED helps in two ways: reducing the effects of non-uniform Bragg disks caused by dynamical scattering or Kikuchi lines, and mitigating issues due to slight off-zone axis alignment. During a PED experiment, the dynamic precession of the electron beam causes the Ewald sphere to rock about the optical axis, cutting through the same 'relrods' (reciprocal lattice rod) multiple times with varying excitation errors. This is akin to a stationary beam condition where the sample precesses around the optic axis[83]. Averaging the NBED patterns during beam precession minimizes the impact of dynamic scattering and results in a diffraction pattern similar to those under purely kinematical conditions[61,84]. Moreover, this precession enhances the capture of high-order reciprocal space reflections, which are more sensitive to the subtle strain variations, and illuminates more Bragg peaks, especially when the sample is off a zone axis[46,62].

A sensitive and efficient pixelated electron detector with an appropriate readout array size is crucial for strain mapping accuracy under NBED mode[43]. DED significantly enhances the SNR of Bragg peaks in NBED patterns at the same exposure time compared to CCD[85], allowing for more accurate detection of Bragg peaks. In addition, for accurate strain mapping based on NBED patterns recorded by a DED, sufficient resolution in diffraction space is required, typically determined by the readout pixel array size when the camera length is fixed. Optimizing the camera length to show at least -4g to 4g Bragg peaks and using a pixel array size of at least 128 × 128 provides a suitable resolution in diffraction patterns for strain mapping. While larger detector arrays (e.g., 1024 × 1024) may improve resolution, they usually result in fewer electrons per pixel per frame at the same beam current, necessitating longer exposure times to maintain SNR, which may be undesirable. A larger detector arrays also leads to challenges related to big data storage and more computational load.

In addition to PED and DED, optimizing the convergence angle of the electron probe can further enhance strain measurement precision. Reducing the convergence angle makes the electron beam more parallel, and reduce the radii of the Bragg disks, effectively reducing Bragg peak overlap. When a Bragg peak shrinks from a disk to a small spot, the detection of Bragg disk center is actually easier as the uniformness of the brightness becomes less critical[35]. However, both using PED and reducing convergence angle increase the probe size, thereby decreasing the spatial resolution in real space. Therefore, a trade-off between strain measurement precision and real-space resolution is inevitable. In this study, we set the convergence semi-angle and the precession angle to 0.8 mrad and 1˚, respectively.

The synergistic effects of precession-assisted 4D-STEM and DED on a FIB-prepared Zr sample with slight local misorientation from [0001] zone axis at room temperature in vacuum is illustrated in **Fig.**

2. The sample is mounted in a gas cell with 30 nm and 40 nm thick $Si_3N_4$ membranes on the bottom heater and top chips, respectively.

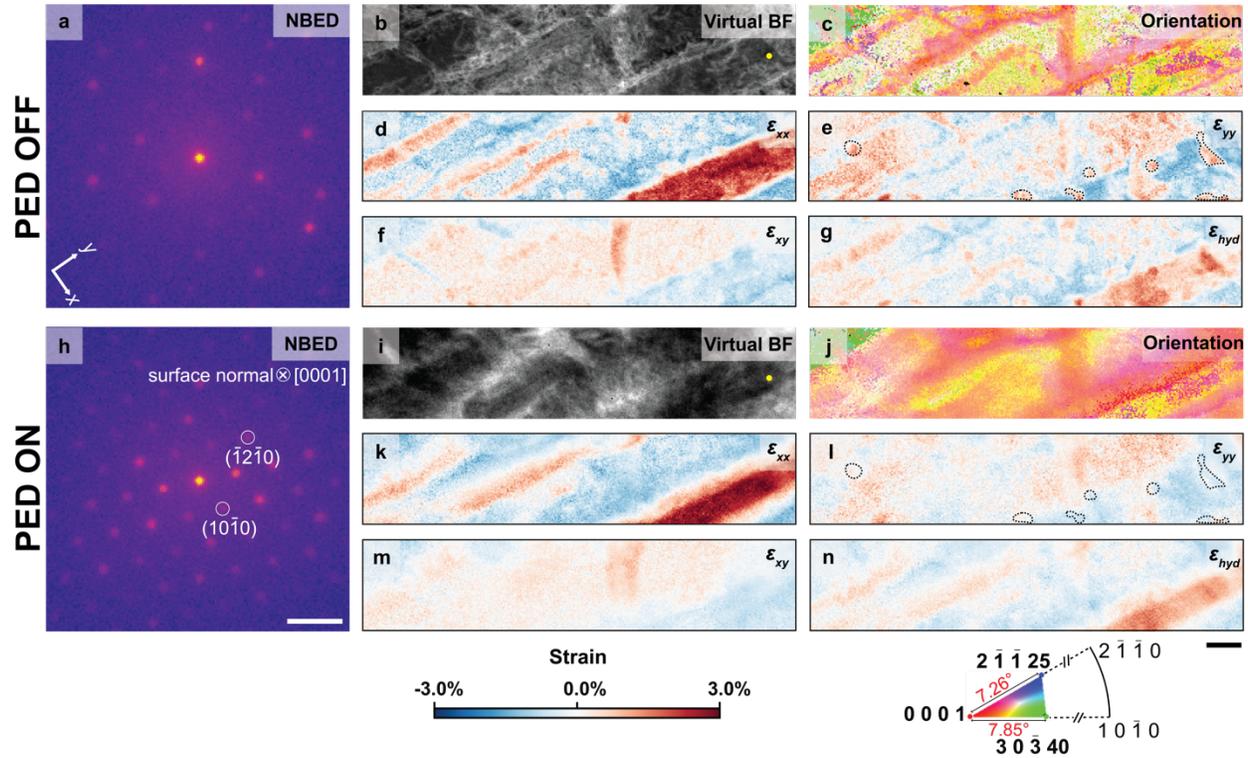

**Fig. 2.** Effects of precession on the quantitative analysis of 4D-STEM data using a pure $\alpha$-Zr sample as an example. (a-g) Representative NBED patterns, virtual BF imaging, strain mapping, and orientation mapping results generated from 4D-STEM dataset recorded without (a-g) and with (h-n) precession. Scale bars are 5 $nm^{-1}$ for diffraction patterns, and 300 nm for real space, respectively. The $[1\,0\,\bar{1}\,0]$ direction of the sample is set as the x-axis for strain mapping calculations, as shown in (a). The falsely detected regions in $\varepsilon_{yy}$ strain mapping are highlighted with black dotted lines in (e), without precession and (l), with precession, respectively.

The NBED pattern in **Fig. 2a** shows a limited number of Bragg spots with non-uniform intensity distribution, likely due to dynamic scattering or an off-zone-axis condition. In contrast, the NBED pattern in **Fig. 2h**, acquired under PED mode at the same location (marked by yellow indicators in the virtual bright-field (BF) images in **Fig. 2b** and **2i**), exhibits enhanced Bragg spot intensity and more closely resembles a kinematic pattern. These differences confirm the effective mitigation of dynamical effects and partial compensation of the inevitable sample misorientation under PED mode.

The virtual BF image (non-PED mode) in **Fig. 2b** appears sharper compared to the BF image in (PED mode) **Fig. 2i**. This difference may origins from several effects of PED, including increased lens aberrations, reduction in delta fringes, and averaging of excitation errors vectors, as summarized in a paper by Rebled et al[86]. First, the precession leads to an increase in probe size and aberrations in the condenser lenses and objective pre-field[44]. Note the spatial resolution is also limited by the choice of step size. In our experiment, the chosen real-space step size in our 4D-STEM experiment is 12 nm, which is similar to the probe size (around 15 nm). The difference of contrast observed in our experiments with and without PED could be mainly attributed to the larger e-beam interaction volume on the sample due to a larger probe size. Second, increased precession angles can diminish the clarity of delta fringes[86]. Third, as the beam precesses, it averages out the excitation error vector, reducing dynamic effects and some diffraction contrast, such bend contours in curved crystals or thickness fringes in the thin foils.

For orientation mapping, our method indexes NBED patterns by matching experimental data to a library of simulated NBED patterns. The orientation maps presented in **Fig. 2c-2j**, are combined with reliability parameter of the method: the higher the reliability value the lighter the colors are, and vice-versa. In current maps, there are many black pixels in the orientation map due to low reliability when precession is turned off, as shown in **Fig. 2c**. These invalid pixels primarily result from the limited number of reflections per pattern when precession is off, and also from strong dynamical scattering. Previous work has indicated that PED enhances the accuracy of orientation mapping[87] by mitigating these issues. Such a beneficial effect of PED is also shown in our result (**Fig. 2j**)[87].

Similar to the orientation mapping, the PED mode results in smoother real-space strain maps, as shown in **Fig. 2d-2g** and **Fig. 2k-2n**, suggesting improved accuracy of strain measurements. Notably, some falsely detected signals in non-PED mode – such as those within the black dotted lines in **Fig. 2e** – are mitigated in PED mode (**Fig. 2l**). This improvement is not merely due to blurring effects. The supplementary **Fig. S1** compares the strain mapping results for PED-off mode without Gaussian blurring, PED-off mode with a Gaussian blurring at various $\sigma$ values, and the PED-on mode without Gaussian blurring results. The comparison illustrates that Gaussian blurring alone cannot eliminate falsely detected signals in strain mapping, suggesting that the enhancement in strain maps by PED is physical.

A quantitative proof of PED's improvement in strain mapping accuracy relies on a comparison with the ground truth, which has been previously discussed and provided in a simulation work by Mahr et al[88].

It is critical to discuss about the effect of out-of-plane sample tilt on the accuracy of strain measurements. To address this, we conducted multi-slice simulation of a strain-free 5 nm thick $\alpha$-Zr sample on [0001] zone axis, with mistilts ranging from 0° to 10°, as illustrated in **Fig. 3**. The simulated SAED patterns were analysed by cross-correlation to identify the Bragg peaks' locations. With increasing out-of-plane tilt, the $(10\bar{1}0)$ and $(1\bar{3}\bar{4}0)$ Bragg peaks not only distort and transform from circular to elliptical shapes gradually, but also shifts gradually. These effects led to the detection of a pseudo-strain, despite the sample being strain-free in our simulation.

To visualize this effect, we overlaid the enlarged $(10\bar{1}0)$ and $(1\bar{3}\bar{4}0)$ Bragg peaks, marking the original Bragg peak locations before tilt (*i.e.*, the ground truth) in red and the detected peak centers under tilt in blue, as shown in **Fig. 3h-j, Fig. 3k-m**. Further, **Fig. 3e** indicated that pseudo-strain measured from low order Bragg peak $(10\bar{1}0)$ remained minimal, not exceeding 0.4 % within the explored tilt range. In contrast, the absolute value of pseudo-strain measured from high order Bragg peak $(1\bar{3}\bar{4}0)$ reached up to 1.5% within the explored tilt range, as shown in **Fig. 3f**. These results suggest that high-order Bragg peaks are more susceptible to tilt effects and are therefore less reliable for strain analysis.

Strain analysis methods based on individual Bragg peak detection, such as **Topspin**[71] and **py4DSTEM**[64], primarily rely on low-order Bragg peaks, which contribute the majority of the signal in the NBED pattern when the sample is close to a low index zone axis and serve as the dominant term in strain analysis. **Fig. 3g** shows that the pseudo-strain found by these algorithms is primarily along the direction perpendicular to the tilt axis ($\varepsilon_{yy}$), which is close to zero until 5° tilt and gradually increases to approximately 1% (compressive, negative sign) at 10° tilt.

Our inverse pole figure in **Fig. 2** shows the misorientation with respect to the [0 0 0 1] direction is roughly smaller than 5°in the ROI, leading to a pseudo-strain up to 0.3%. Therefore, our strain measurements are reliable within the scope of this study.

It is worth noting that the strain variation in **Fig. 2k-n** may be attributed to the hydrides formed locally during the FIB sample preparation process[89,90].

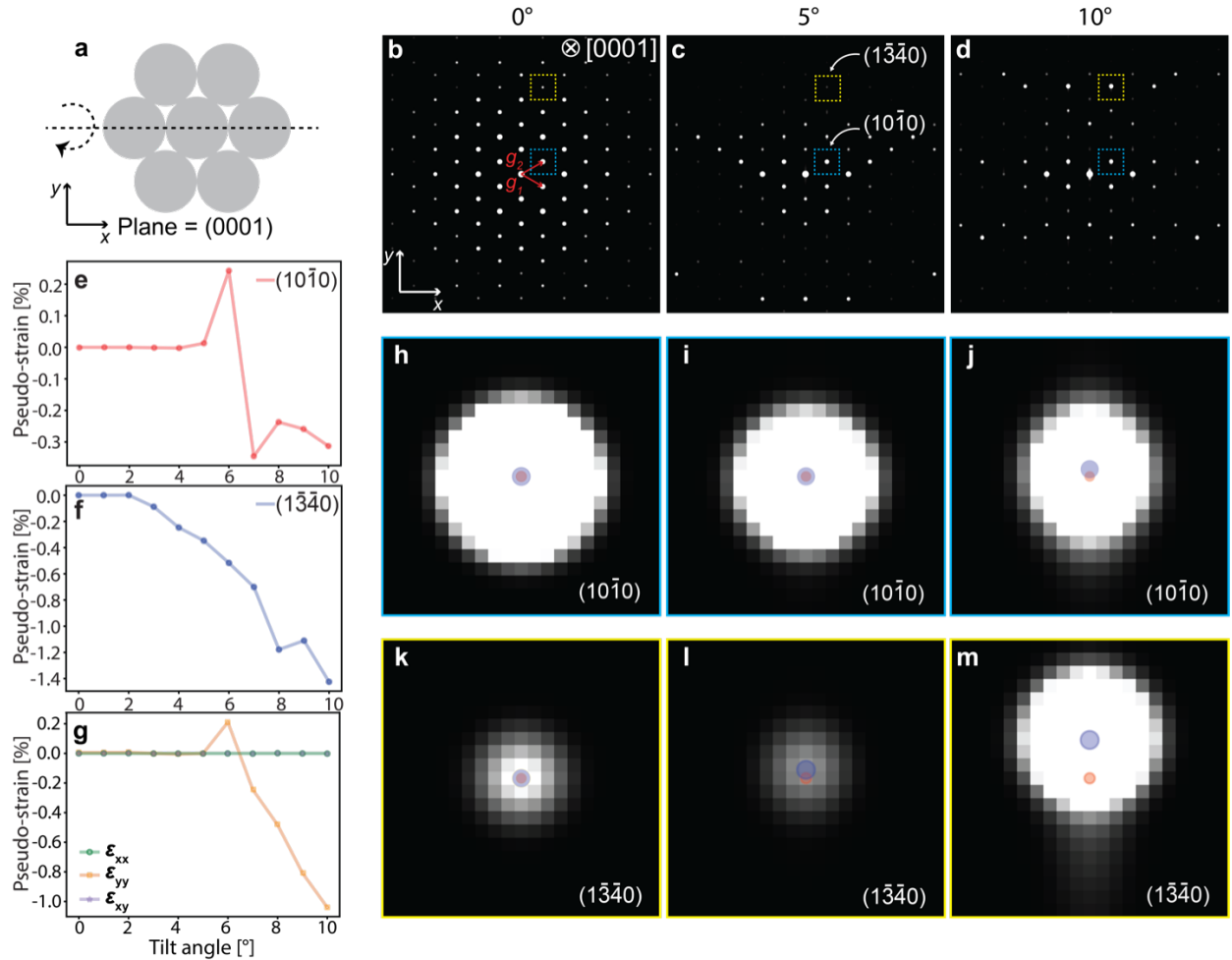

**Fig. 3.** Multi-slice simulation results illustrating the influence of sample tilt on strain analysis based on NBED. (a) Schematic representation of the sample tilt direction. (b-d) Simulated diffraction patterns for a strain-free 5 nm thick α-Zr with increasing mistilts of 0°, 5° and 10°. At 0°, the sample surface normal is aligned with [0001] direction. (e-f) Pseudo-strain measured using (1 0 $\bar{1}$ 0) Bragg peaks and (1 $\bar{3}$ $\bar{4}$ 0) Bragg peaks, respectively. (g) Pseudo-strain analyzed by the py4DSTEM strain mapping module. (h-j) Enlarged view of the low-order (1 0 $\bar{1}$ 0) Bragg peak, corresponding to the yellow-boxed regions in (b-d). (k-m) Enlarged view of the high-order (1 $\bar{3}$ $\bar{4}$ 0) Bragg peak, corresponding to the blue-boxed regions in (b-d). The red spots in (h-m) indicate the original Bragg peak centers at 0° tilt (*i.e.,* the ground truth), while the blue spots represent their detected positions after tilting.

*3.3 Precession-assisted 4D-STEM in gas environments*

In a gas-cell experiment, the extra electron scattering from the *in situ* TEM gas cell window material and the enclosed gas not only makes the electron beam less coherent but also enhance the chances for multiple scattering and inelastic scattering, ultimately degrading the SNR and the spatial resolution in electron microscopy (EM) images[91,92].

Our discussion above (section 3.2) highlights that integrating PED and DED can effectively enhance orientation and strain analysis when the sample is in a gas cell under vacuum conditions -- both internally and externally. In this section, we will consider the scenario when the gas is introduced into the gas cell.

Typically, high gas pressure is necessary to replicate the material's working conditions[93,94], while lower gas pressure is preferred for reducing extra scattering, thus achieving higher SNR and spatial resolution for imaging. To find an optimal balance between these two considerations, we performed a series of precession-assisted 4D-STEM at various Ar gas pressures, including 0 mbar, 300 mbar, 500 mbar, and 1000 mbar. These results are summarized in **Fig. 4**.

As shown in the NBED column of **Fig. 4a-d**, the increasing internal gas pressure in the gas cell results in more and more noticeable deterioration in the SNR. Consequently, both the number of detected Bragg peaks (BPs) and the $|k|_{max}$ (defined as the maximum distance between a detected BP and the transmitted beam, where $k$ is the reciprocal lattice vector) decrease as the gas pressure increases (**Fig. 4e-f**). The yellow dots overlaid on the virtual BF images in **Fig. 4a-d** indicate the locations where the representative NBED patterns were obtained. Furthermore, the orientation and strain maps at higher gas pressures exhibit not only lower spatial resolution but also more false-detected pixels and even invalid pixels (*i.e.*, pixels that could not be analyzed by the code, marked in black color). As shown in **Fig. 4g**, the number of invalid pixels increases dramatically when the gas pressure exceeds 300 mbar.

Our quantitative analysis highlights the importance of controlling gas pressure to enable reliable and precise 4D-STEM characterization. While balancing the optimal pressure for reactions with that for imaging can be challenging, our study suggests that starting with a gas pressure of no more than 300 mbar could be effective, particularly for our relatively thick sample (100 - 200 nm).

The gas inside the cell generates a significant number of inelastically scattered electrons, creating a diffuse background in the NBED pattern (**Fig. 4**). To further improve the accuracy of strain mapping, placing a DED after an energy filter could be beneficial, as the energy filter effectively removes these electrons, enhancing the SNR in NBEDs[41,95]. Additional improvements can be achieved by further reducing the thickness of the $Si_3N_4$ gas cell window material.

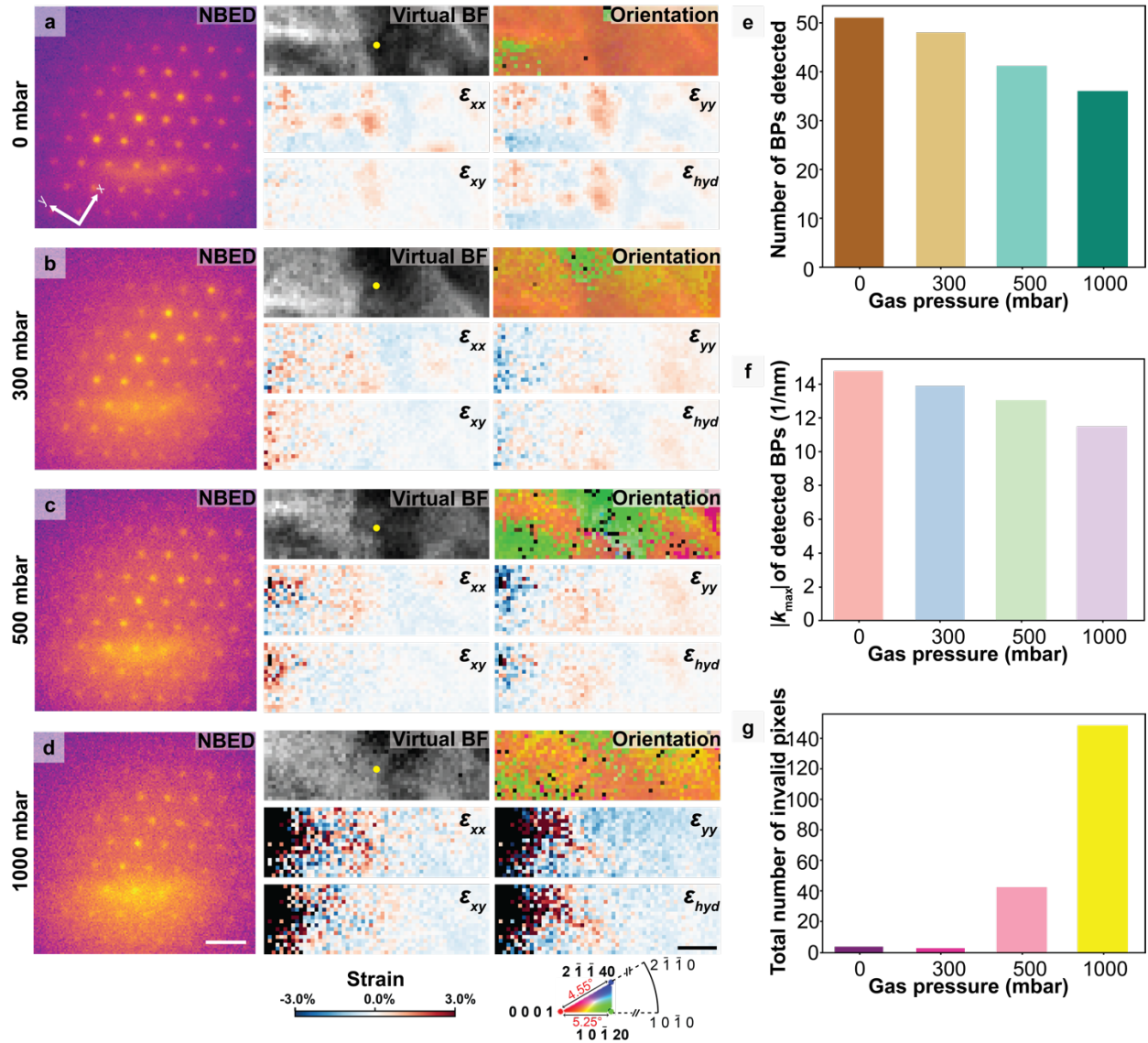

**Fig. 4.** Effects of gas pressure on the quantitative analysis of precession-assisted 4D-STEM data. (a-d) Representative NBED patterns, virtual imaging, strain mapping, and orientation mapping results generated from 4D-STEM datasets recorded at various argon gas pressures, including 0 mbar, 300 mbar, 500 mbar, and 1000 mbar. Invalid pixels are shown in black. Scale bars are 5 nm$^{-1}$ for diffraction patterns, and 300 nm for real-space images. The $[1\,0\,\bar{1}\,0]$ direction of the sample is set as the x-axis for strain mapping calculations, as shown in (a). (e) Average number of Bragg peaks (BPs) detected in the NBED patterns at different gas pressures. (f) Average $|k|_{max}$ of detected BPs in the NBED patterns at different gas pressures. (g) Total number of invalid pixels in the strain and orientation mappings at different gas pressures.

### 3.4. Pausing material evolution for reliable 4D-STEM data acquisition

*In situ* observation of dynamic processes requires an imaging frame rate much shorter than the timescale of significant material changes. Otherwise, captured signals may correspond to an average of multiple unique materials' states over time, making each frame unrepresentative of a specific state at a specific moment. Recording a 4D-STEM dataset typically takes several minutes[60,98], especially when the scan size (*i.e.,* the number of points in the probe position grid) is large, or long-exposure NBED patterns are required.

Despite advances in pixelated direct electron detectors, capturing a single 4D-STEM scan, consisting of thousands to millions of nanobeam electron diffraction patterns, can take anywhere from several seconds to several minutes. If there is not enough electrons (when sample is too thick or too thin) or additional factors such as system drift occur, even longer acquisition times may be needed[97]. Since some chemical reactions can complete in under a minute, a practical solution is to temporarily pause the reaction during data collection and resume it afterward. This approach ensures that each 4D-STEM scan accurately represents a specific time point.

For *in situ* gas cell experiments at high temperatures, there are two potential approaches to pause the reaction. The first involves temperature quenching, where the reaction temperature is reduced to a significantly lower temperature to slow or stop the reaction. This method provides a rapid response but leaves gas in the gas-cell, leading to a lowers imaging quality. The second approach alters the gas supply by replacing the reactive gas (e.g., oxygen) with an inert gas (e.g., argon) or evacuating the gas cell to vacuum. While this may improve imaging quality, especially in a vacuum, it can take a much longer time to halt the reaction, as the gas pressure control is much less responsive than the temperature, and the residual reactive species inside or on the surface of the sample (such as hydrogen and oxygen) may continue reacting with the sample even after the internal pressure of the gas-cell has reached its best vacuum state.

To overcome these limitations, we combine both methods using the DENSsolutions Climate system, which applies a series of temperature and pressure pulses. This approach ensures rapid reaction pausing while improving the imaging quality.

**Fig. 5a** illustrates the preset temperature $T_{set}$ and pressure $P_{set}$ profiles in our experiments, which comprise sequential composite pulses of heating and pressure. **Fig. 5b** shows a representative composite pulse, featuring a shorter temperature pulse superimposed on a longer gas flow pulse. The start of the pressure pulse precedes that of the temperature pulse to compensate for the slower response of the gas pressure. To pause the reaction, we terminate the gas supply and allow the sample to cool naturally to room temperature (25°C). At the end of each composite pulse, both heating and gas pressure are simultaneously deactivated. The effective reaction period and the 4D-STEM data collection period are indicated in **Fig. 5b**.

The measured profiles for pressure ($P_{Measured}$), gas flow rate ($F_{Measured}$), and temperature ($T_{Measured}$) in the nano-reactor are shown in **Fig. 5c**. Although we set the $P$ and $T$ profiles for rapid changes (like a step-function), the measured profiles differ significantly from the preset. Notably, there is considerable fluctuation at the beginning of the $P_{Measured}$ and $F_{Measured}$ pulses, likely due to hardware constraints. Thus, it is crucial to raise the temperature and start the reaction only after the $P$ and $F$ stabilize. It takes around 66.9 seconds for $P$ to stabilize during the pressure ramping stage. The pressure decrease is slower, requiring around 100 seconds to drop to 3% of its initial value before pumping ($P^*$). The decay rate diminishes at lower pressures, making it inefficient to pause the reaction solely by deactivating the pressure due to the prolonged time needed to re-establish a strong vacuum. In contrast, temperature control is more prompt, taking 5.6 seconds to ramp the temperature to 350 °C (an average temperature ramping rate of ~ 58.0 °C/s), and 4.8 seconds to cool down to 32 °C naturally. The cooling profile also has a long tail because the driving force for heat dissipation (*i.e.,* temperature difference) reduces as the temperature approaches to the ambient temperature of the surrounding components (chip, holder, TEM column), maintained at 25°C.

To assess the impact of the slow temperature response on the effective reaction time, we performed the following analysis. In the preset temperature profile, the sample was maintained at 350°C for 10 seconds, with rapid transitions from and to 25°C (**Fig. 5b**). However, $T_{Measured}$ indicates that the sample remained above 348°C for only 3.4 seconds, above 300°C for 8.06 seconds, and above 26°C for 17.64 seconds (**Fig. 5c**) during a single temperature pulse. Understanding the deviation between the actual and preset temperature profiles is critical for the reliable design of gas-cell *in situ* TEM experiments and for accurate analysis of the effective reaction time.

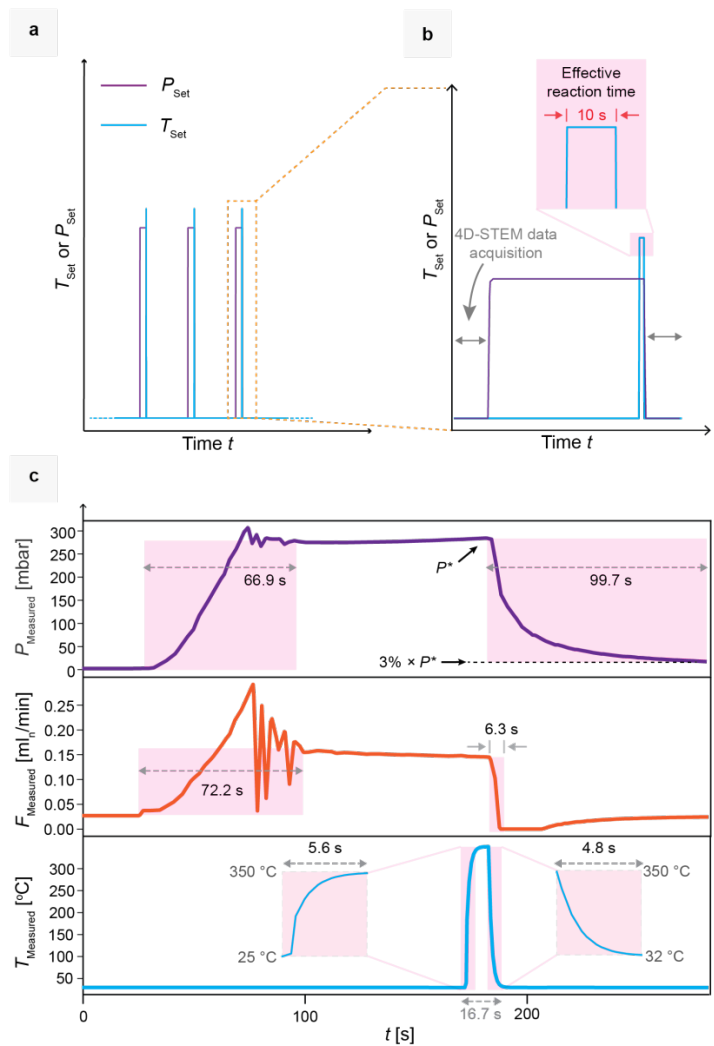

**Fig. 5.** Pressure and temperature profiles for pausing material evolution to enable reliable 4D-STEM data collection. (a) Preset pressure and temperature profile with three sequential composite pulses. (b) Enlarged view of a single preset pressure-temperature composite pulse. The inset shows a detailed view of the temperature pulse. (c) The measured profiles for pressure ($P_{Measured}$), gas flow rate ($F_{Measured}$), and temperature ($T_{Measured}$) in the nano-reactor during a composite pulse.

*3.5 In situ precession-assisted 4D-STEM analysis of strain and orientation evolution during zirconium oxidation*

The oxidation of pure zirconium has been chosen to demonstrate the effectiveness of our methods for strain and orientation mapping with precession-assisted *in situ* 4D-STEM.

Zirconium-based alloys are commonly used as cladding materials for encapsulating nuclear fuel pellets in light water reactors, which typically operate at temperatures between 288 °C and 330 °C[7]. The zirconium oxidation process induces stress in both the substrate and oxide layers due to volumetric expansion as the metal transforms into oxide, characterized by a Pilling-Bedworth ratio of 1.56[98]. This stress buildup and anisotropic strain accumulation are hypothesized to contribute to the formation of lateral cracks and pores, which serve as short-circuit paths[99] for oxygen diffusion to the oxide/metal interface, thereby accelerating corrosion. While cracks and pores are inevitable during oxidation, the "breakaway" phenomenon in zirconium oxidation is more catastrophic and behaves as a critical transition point: the formation of a percolated network of pores and cracks in the oxides leads to a significant increase in oxidation rate. Following breakaway, a new oxidation cycle begins, eventually leading to another breakaway event. As a result, the oxidation of zirconium and many of its alloys follows a cyclic kinetics, posing significant challenges to the safety and economic efficiency of light water nuclear reactors.

This study of pure zirconium oxidation is focused on the initial stage and thin films, thus the oxidation mechanisms in our experiments could be significantly different from that in bulk scale and nuclear-relevant environment. Specifically, the "breakaway" phenomenon[100–102] is absent due to the limitation of experimental time- and length- scales. However, understanding the nanoscale strain/orientation evolution during the initial oxidation[103] of metals is meaningful for developing a more comprehensive theory of metal oxidation and providing insights into new alloys development.

Previous ex-situ studies on Zr oxidation, such as transmission electron backscatter diffraction (t-EBSD)[104,105], automated crystal orientation mapping[106,107] and scanning precession electron diffraction (SPED)[108], have been utilized to provide the grain-to-grain misorientation mapping for studying oxide microstructure evolution and tetragonal-$ZrO_2$ to monoclinic-$ZrO_2$ transformation during the oxidation of zirconium and its alloys. More recently, *in situ* TEM[58,109,110] study of zirconium oxidation using a gas-cell holder combined with ex-situ SPED was also reported[58]. While these studies have provided valuable information about morphological and orientation changes during oxidation, to our best knowledge, *in situ* TEM characterization of strain evolution in Zr metal during its oxidation has not been reported in the open literature. Therefore, we conducted *in situ* strain mapping using our technique to study the initial oxidation of pure zirconium at 350 °C.

**Fig. 6a** presents hydrostatic strain and orientation mapping of the α-Zr substrate before and after 100 seconds of oxidation, as per the preset profile described in section 3.4. Each NBED pattern captures structural data for the zirconium oxides and the α-Zr metal substrate, with drift correction ensuring accurate strain and orientation mapping comparisons.

In the early stages of oxidation, when the zirconium metal surface interacts with oxygen, two key processes occur. First, oxygen is incorporated into the Zr metal lattice as interstitials[111], forming a solid solution (SS)[112]. Second, at higher oxygen concentrations, suboxide[104,113,114] or di-oxide phases begin to form. Both processes result in volumetric expansion within the surface layer that interacts with oxygen (*i.e.*, the $ZrO_x$ layer). This expansion induces tensile stress in the underlying unreacted metal, as schematically illustrated in **Fig. 6b**. However, it is critical to note that the first process (*i.e.,* forming SS) mainly leads to expansion along the *c*-axis, while its effect on the lattice spacing along *a*-axis is ignorable[112], only up to 0.6%. In our study, the average in-plane hydrostatic strain (*i.e.*, tensile strain along *a*-axis) in the α-Zr metal phase increased from 0 to 2.58% post-oxidation (**Fig. 6a**), which should be mainly attributed to the formation of oxide at the metal surface. Additionally, the hydrostatic strain map exhibits local strain 'hotspots,' indicating heterogeneous strain evolution. As tensile strain tends to facilitate crack initiation and propagation, these hotspots may act as precursors to nano cracks. The evolution and localization of elastic strain can potentially influence the local roughness of surfaces and interfaces, porosity of oxide and metals, and undulation of the metal-oxide interface, thereby impacting oxidation kinetics.

Before oxidation, the α-Zr substrate aligns near the [0001] zone axis, although slight misorientation is observed due to local bending, defects or hydrides, as shown in **Fig. 6a** and the detailed analysis in **Fig. 7**. Post-oxidation, this misorientation intensifies and becomes more heterogeneous, as the formation of oxide phases induces local bending, strain or grain rotation.

This section showcases how *in situ* precession-assisted 4D-STEM provides spatially resolved strain and orientation mapping, offering critical insights into the evolution of strain localization and crystal orientation during chemomechanical processes at the nanoscale.

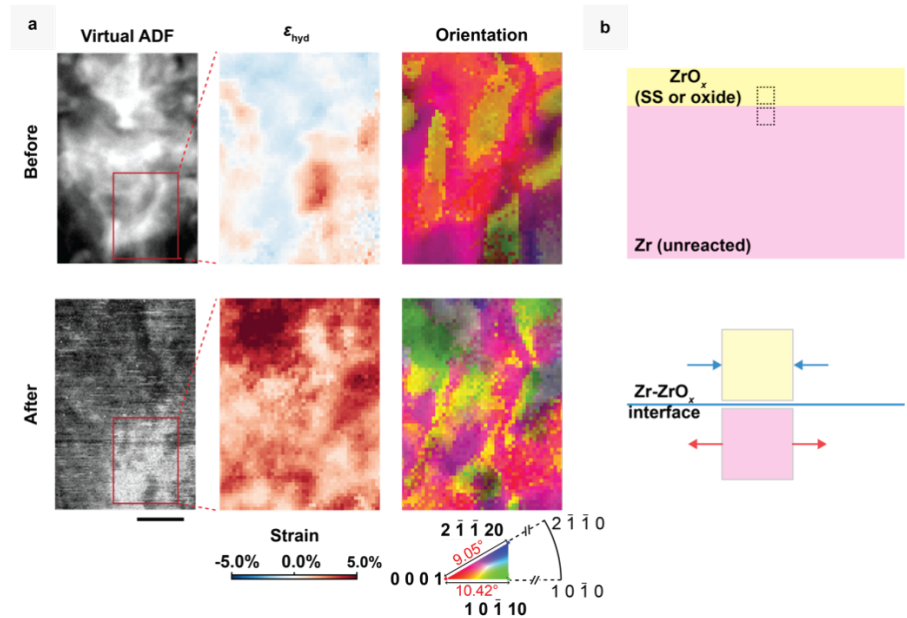

**Fig. 6** *In situ* precession-assisted 4D-STEM analysis of pure zirconium oxidation. (a) Virtual annular dark-field (ADF) imaging, hydrostatic strain mapping, and orientation mapping of the sample before and after 100 seconds of oxidation. Scale bar, 300 nm. (b) Schematic illustration depicting strain evolution near the Zr-ZrO$_x$ interface. Volume expansion in the surface ZrO$_x$ layer induces tensile strain in the Zr metal substrate and compressive strain in the oxide layer.

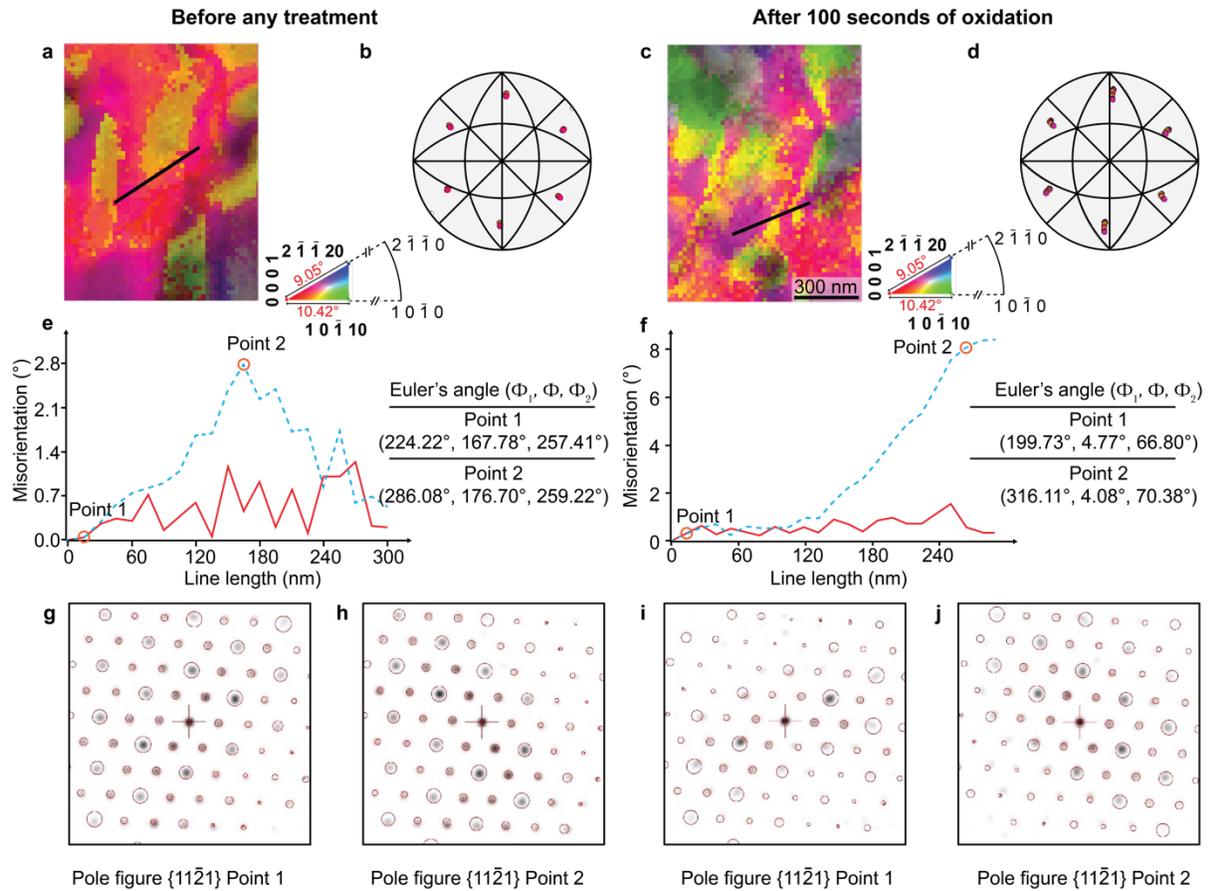

**Fig. 7** Misorientation analysis for sample before and after 100 seconds of oxidation. (a, c) Local orientation map shown in **Fig. 6**, with a black line drawn for misorientation line profile analysis. (b, d) Pole figure {11$\bar{2}$1} along the black line in the orientation map. (e, f) Misorientation line profile along the black lines in (a) and (c). The red line indicates the misorientation from one pixel to the next along the line, while the blue lines indicate the misorientation of each pixel relative to the first point of the line. (g, h, i, j) Experimental diffraction patterns overlaid with matching simulated patterns for points 1 and 2 in (e) and (f).

## 4. Conclusion

    A comprehensive understanding of strain and orientation evolution in materials during reactions in gas or liquid environments is crucial for elucidating the chemomechanical interactions in advanced structural and functional materials under operational conditions. Such an understanding is pivotal for further enhancing the materials' performance or better predicting their failure. In this study, we developed *in-situ* precession-assisted 4D-STEM experiments with a practical sample preparation procedure to generate nanoscale strain and orientation mapping for gas cell experiments. Compared to other previous methods, such as GPA or EBSD, our method provides a suitable balance of spatial resolution and FOV.

    Using this method, we studied the strain and orientation evolution during pure Zr oxidation. The use of PED and DED in the 4D-STEM experiments effectively eliminates dynamical effects and increases the SNR in NBED patterns. Consequently, the reduced non-uniformity and increased detectable numbers

of Bragg peaks in reciprocal space enhance the accuracy and robustness of orientation and strain mapping measurements. Furthermore, our quantitative analysis of the effects of gas pressure demonstrates that optimizing gas pressure is vital for minimizing 4D-STEM imaging imperfections. By designing a pressure and temperature profile to pause material evolution, reliable precession-assisted 4D-STEM data collection is enabled.

Additionally, we investigated the impact of out-of-plane sample tilt on strain measurement accuracy using multislice simulations. Cross-correlation analysis of simulated SAED patterns revealed that increasing tilt distorts and shifts Bragg peaks, producing a pseudo-strain despite the absence of real strain. Low-order Bragg peaks exhibited minimal pseudo-strain (<0.4%), whereas high-order peaks like showed larger deviations (up to 1.5%), indicating greater sensitivity to tilt. Since most strain analysis codes rely on low-order peaks, the resulting strain artifacts were generally limited (<1%) if the low order peaks are bright enough (typically can be facilitated by PED).

To validate the effectiveness of our workflow, we applied it to study the early-stage oxidation of Zr. The results show that the Zr metal experiences tensile strain following initial oxidation, consistent with the well-established theory that oxidation leads to strain at the metal–oxide interface, with tension on the metal side and compression on the oxide side. This agreement demonstrates the reliability of our method.

While our case study is based on Zr initial oxidation, our method can also be applied to other material systems such as catalysts[8,9], batteries[3–6], or other structural materials working in extreme environments[115] to facilitate the understanding of chemomechanics that impact their performance or degradation.

Looking ahead, several future directions could further enhance our methods.

First, sample preparation would benefit from a FIB-free ex-situ transfer process[116,117]. The Ga/Pt/C contamination[118–120] caused by *in situ* FIB transfer during welding or imaging can complicate the reactions[82], making it more challenging to decipher the fundamental mechanisms. Incorporating cryogenic FIB and the EXpressLO™ ex-situ transfer techniques[58,121] can help reduce hydrides, leading to samples with less artefacts or contaminations and a clearer understanding of zirconium degradation mechanism.

Second, hardware improvements could further enhance strain mapping accuracy, such as using an energy filter to eliminate inelastic scattered electrons and improve the SNR of NBED[41], integrating a Bullseye aperture to enhance Bragg Peak detection[42,46], employing a gas cell with thinner $Si_3N_4$ window membrane[122], and increasing the beam current if the sample is not beam-sensitive.

Third, the accuracy of strain mapping can depend considerably on the software used and the appropriateness of spot detection settings. There are a variety of choices of diffraction-based strain analysis software or codes now, such as Topspin[71], py4DSTEM[64], and Pyxem[123] etc. While many of them are based on individual Bragg peak detection and cross-correlation, their intrinsic algorithm could be significantly different, such as whether Radon transformation is used or weather there is support for multi-core CPU and GPU acceleration. Recently, there are also new algorithms that leverages machine learning or deep learning[124–127]. For example, FCU-Net deep learning methods[128] has shown some advantages over the cross-correlation method. On the other hand, a "Whole Pattern Fitting" method[129] is developed to overcome the challenge of strain analysis on diffraction patterns with multiple overlapping lattices.

Last but not least, more responsive temperature and gas adjustments for reaction control that reduces stabilization time, would improve experimental efficiency and enable better control over effective reaction durations.

By integrating these advancements, *in situ* gas phase precession-assisted 4D-STEM can be extended to a broader range of material systems and experimental conditions, enhancing characterization accuracy and providing deeper insights into the nanoscale evolution of materials under operational conditions.

**Author Contribution**

Y.Y. and D.Z. conceived the idea and designed the experiments. Y.H. prepared the FIB sample. A.S.G., A.G.P., and D.Z. performed the precession-assisted 4D-STEM experiments. Y.S., Y.Y., A.S.G., A.G.P., K.W. and D.Z. analyzed the data and discussed the results. Y.S., Y.Y., and D.Z. wrote the manuscript. Y.Y., D.Z., S.N, and H.P.G secured funding for the project. All authors contributed to the manuscript preparation and discussion of the results.

**Declaration of Competing Interest**

All authors declare no conflict of interest in this manuscript.

**Acknowledgements**

This research is supported by the National Science Foundation (NSF) early career award DMR-2145455. Financial support by the European Union's HORIZON 2020 Research and Innovation Programme ESTEEM3 under grant agreement No. 823717 is gratefully acknowledged.

**Reference**

1. De Vasconcelos, L. S. *et al.* Chemomechanics of Rechargeable Batteries: Status, Theories, and Perspectives. *Chem. Rev.* **122**, 13043–13107 (2022).
2. Yildiz, B., Nikiforova, A. & Yip, S. Metallic interfaces in harsh chemo-mechanical environments. *Nucl. Eng. Technol.* **41**, 21–38 (2009).
3. Yan, P. *et al.* Intragranular cracking as a critical barrier for high-voltage usage of layer-structured cathode for lithium-ion batteries. *Nat. Commun.* **8**, 1–9 (2017).
4. McDowell, M. T., Quintero Cortes, F. J., Thenuwara, A. C. & Lewis, J. A. Toward high-capacity battery anode materials: Chemistry and mechanics intertwined‖. *Chem. Mater.* **32**, 8755–8771 (2020).
5. Salah, M. *et al.* Pure silicon thin-film anodes for lithium-ion batteries: A review. *J. Power Sources* **414**, 48–67 (2019).
6. Zhao, Y. *et al.* A review on modeling of electro-chemo-mechanics in lithium-ion batteries. *J. Power Sources* **413**, 259–283 (2019).
7. Motta, A. T., Couet, A. & Comstock, R. J. Corrosion of Zirconium Alloys Used for Nuclear Fuel Cladding. *Annu. Rev. Mater. Res.* **45**, 311–343 (2015).
8. Strasser, P. *et al.* Lattice-strain control of the activity in dealloyed core-shell fuel cell catalysts. *Nat. Chem.* **2**, 454–460 (2010).
9. Feng, Q. *et al.* Strain Engineering to Enhance the Electrooxidation Performance of Atomic-Layer Pt on Intermetallic $Pt_3Ga$. *J. Am. Chem. Soc.* **140**, 2773–2776 (2018).
10. Kim, S. *et al.* Electrochemically driven mechanical energy harvesting. *Nat. Commun.* **7**, (2016).
11. Yildirim, C., Cook, P., Detlefs, C., Simons, H. & Poulsen, H. F. Probing nanoscale structure and

12. Richter, C. *et al.* Nanoscale Mapping of the Full Strain Tensor, Rotation, and Composition in Partially Relaxed InxGa1-x N Layers by Scanning X-ray Diffraction Microscopy. *Phys. Rev. Appl.* **18**, (2022).
13. Hofmann, F. *et al.* Nanoscale imaging of the full strain tensor of specific dislocations extracted from a bulk sample. *Phys. Rev. Mater.* **4**, 013801 (2020).
14. Krawitz, A. D. & Holden, T. M. The Measurement of Residual Stresses Using Neutron Diffraction. *MRS Bull.* **15**, 57–64 (1990).
15. Plotkowski, A. *et al.* Operando neutron diffraction reveals mechanisms for controlled strain evolution in 3D printing. *Nat. Commun.* **14**, 1–11 (2023).
16. Wright, S. I., Kacher, J. & Ruggles, T. Electron Backscatter Diffraction based Strain Analysis in the Scanning Electron Microscope. 1–31 (2021).
17. Wright, S. I., Nowell, M. M. & Field, D. P. A Review of Strain Analysis Using Electron Backscatter Diffraction. *Microsc. Microanal.* **17**, 316–329 (2011).
18. Dombrowski, K. F. & De Wolf, I. Stress measurements in sub-μm Si structures using raman spectroscopy. *Solid State Phenom.* **63–64**, 519–524 (1998).
19. Im, H. S. *et al.* Strain Mapping and Raman Spectroscopy of Bent GaP and GaAs Nanowires. *ACS Omega* **3**, 3129–3135 (2018).
20. Naresh-Kumar, G. *et al.* Non-destructive imaging of residual strains in GaN and their effect on optical and electrical properties using correlative light-electron microscopy. *J. Appl. Phys.* **131**, (2022).
21. Hÿtch, M. J. & Minor, A. M. Observing and measuring strain in nanostructures and devices with transmission electron microscopy. *MRS Bull.* **39**, 138–146 (2014).
22. Han, Y. *et al.* In Situ TEM Characterization and Modulation for Phase Engineering of Nanomaterials. *Chem. Rev.* (2023) doi:10.1021/acs.chemrev.3c00510.
23. An, Z. *et al.* Negative mixing enthalpy solid solutions deliver high strength and ductility. *Nature* **625**, 697–702 (2024).
24. Amini, S. *et al.* In operando 3D mapping of elastic deformation fields in crystalline solids. *Matter* **7**, 2591–2608 (2024).
25. Vatanparast, M. *et al.* Strategy for reliable strain measurement in InAs/GaAs materials from high-resolution Z-contrast STEM images. *J. Phys. Conf. Ser.* **902**, (2017).
26. Zhu, Y., Ophus, C., Ciston, J. & Wang, H. Interface lattice displacement measurement to 1 pm by geometric phase analysis on aberration-corrected HAADF STEM images. *Acta Mater.* **61**, 5646–5663 (2013).
27. Hÿtch, M. J., Snoeck, E. & Kilaas, R. Quantitative measurement of displacement and strain fields from HREM micrographs. *Ultramicroscopy* **74**, 131–146 (1998).
28. Chung, J. & Rabenberg, L. Effects of strain gradients on strain measurements using geometrical phase analysis in the transmission electron microscope. *Ultramicroscopy* **108**, 1595–1602 (2008).
29. Bierwolf, R. *et al.* Direct measurement of local lattice distortions in strained layer structures by HREM. *Ultramicroscopy* **49**, 273–285 (1993).
30. Luo, X. *et al.* High-precision atomic-scale strain mapping of nanoparticles from STEM images. *Ultramicroscopy* **239**, 113561 (2022).
31. Nord, M., Vullum, P. E., MacLaren, I., Tybell, T. & Holmestad, R. Atomap: a new software tool for the automated analysis of atomic resolution images using two-dimensional Gaussian fitting. *Adv. Struct. Chem. Imaging* **3**, (2017).
32. Hÿtch, M. J. & Plamann, T. Imaging conditions for reliable measurement of displacement and strain in high-resolution electron microscopy. *Ultramicroscopy* **87**, 199–212 (2001).
33. Braidy, N., Le Bouar, Y., Lazar, S. & Ricolleau, C. Correcting scanning instabilities from images of periodic structures. *Ultramicroscopy* **118**, 67–76 (2012).
34. Jones, L. & Nellist, P. D. Identifying and correcting scan noise and drift in the scanning transmission electron microscope. *Microsc. Microanal.* **19**, 1050–1060 (2013).

35. Rouviere, J. L., Béché, A., Martin, Y., Denneulin, T. & Cooper, D. Improved strain precision with high spatial resolution using nanobeam precession electron diffraction. *Appl. Phys. Lett.* **103**, (2013).
36. Cooper, D. & Rouviere, J. L. Strain measurement with nanometre resolution by transmission electron microscopy. *Adv. Mater. Res.* **996**, 3–7 (2014).
37. Clément, L., Pantel, R., Kwakman, L. F. T. & Rouvière, J. L. Strain measurements by convergent-beam electron diffraction: The importance of stress relaxation in lamella preparations. *Appl. Phys. Lett.* **85**, 651–653 (2004).
38. Houdellier, F., Roucau, C., Clément, L., Rouvière, J. L. & Casanove, M. J. Quantitative analysis of HOLZ line splitting in CBED patterns of epitaxially strained layers. *Ultramicroscopy* **106**, 951–959 (2006).
39. Cooper, D. *et al.* Strain mapping with nm-scale resolution for the silicon-on-insulator generation of semiconductor devices by advanced electron microscopy. *J. Appl. Phys.* **112**, (2012).
40. Béché, A., Rouvière, J. L., Barnes, J. P. & Cooper, D. Strain measurement at the nanoscale: Comparison between convergent beam electron diffraction, nano-beam electron diffraction, high resolution imaging and dark field electron holography. *Ultramicroscopy* **131**, 10–23 (2013).
41. Yang, Y. *et al.* One dimensional wormhole corrosion in metals. *Nat. Commun.* **14**, (2023).
42. Yang, Y. *et al.* Rejuvenation as the origin of planar defects in the CrCoNi medium entropy alloy. *Nat. Commun.* **15**, 1402 (2024).
43. Ophus, C. Four-Dimensional Scanning Transmission Electron Microscopy (4D-STEM): From Scanning Nanodiffraction to Ptychography and Beyond. *Microsc. Microanal.* 563–582 (2019) doi:10.1017/S1431927619000497.
44. Ozdol, V. B. *et al.* Strain mapping at nanometer resolution using advanced nano-beam electron diffraction. *Appl. Phys. Lett.* **106**, (2015).
45. Cooper, D., Denneulin, T., Bernier, N., Béché, A. & Rouvière, J. L. Strain mapping of semiconductor specimens with nm-scale resolution in a transmission electron microscope. *Micron* **80**, 145–165 (2016).
46. Zeltmann, S. E. *et al.* Patterned probes for high precision 4D-STEM bragg measurements. *Ultramicroscopy* **209**, 112890 (2020).
47. THOMAS, G. *Kikuchi Electron Diffraction and Applications. Diffraction and Imaging Techniques in Material Science* (NORTH-HOLLAND PUBLISHING COMPANY, 1978). doi:10.1016/b978-0-444-85128-4.50015-1.
48. Cowley, J. M. & Moodie, A. F. The scattering of electrons by atoms and crystals. I. A new theoretical approach. *Acta Crystallogr.* **10**, 609–619 (1957).
49. Cheng, Y. *et al.* Understanding all solid-state lithium batteries through in situ transmission electron microscopy. *Mater. Today* **42**, 137–161 (2021).
50. Zhang, L. *et al.* Lithium whisker growth and stress generation in an in situ atomic force microscope–environmental transmission electron microscope set-up. *Nat. Nanotechnol.* **15**, 94–98 (2020).
51. Zhang, Y., Li, Y., Wang, Z. & Zhao, K. Lithiation of SiO2 in Li-ion batteries: In situ transmission electron microscopy experiments and theoretical studies. *Nano Lett.* **14**, 7161–7170 (2014).
52. Wang, P. *et al.* Electro–Chemo–Mechanical Issues at the Interfaces in Solid-State Lithium Metal Batteries. *Adv. Funct. Mater.* **29**, (2019).
53. Hansen, T. W. & Wagner, J. B. Catalysts under controlled atmospheres in the transmission electron microscope. *ACS Catal.* **4**, 1673–1685 (2014).
54. Wu, F. & Yao, N. Advances in windowed gas cells for in-situ TEM studies. *Nano Energy* **13**, 735–756 (2015).
55. Pérez Garza, H. H. *et al.* MEMS-based system for in-situ biasing and heating solutions inside the TEM. in *European Microscopy Congress 2016: Proceedings* 237–238 (Wiley-VCH Verlag GmbH & Co. KGaA, 2016). doi:10.1002/9783527808465.EMC2016.6710.
56. Allard, L. F. *et al.* Novel MEMS-Based Gas-Cell/Heating Specimen Holder Provides Advanced

Imaging Capabilities for In Situ Reaction Studies. *Microsc. Microanal.* **18**, 656–666 (2012).
57. Allard, L. F. *et al.* A new MEMS-based system for ultra-high-resolution imaging at elevated temperatures. *Microsc. Res. Tech.* **72**, 208–215 (2009).
58. Harlow, W., Ghassemi, H. & Taheri, M. L. Determination of the initial oxidation behavior of Zircaloy-4 by in-situ TEM. *J. Nucl. Mater.* **474**, 126–133 (2016).
59. Foucher, A. C. & Stach, E. A. High Pressure Transmission Electron Microscopy (TEM). in *Springer Handbooks* 381–407 (2023). doi:10.1007/978-3-031-07125-6_19.
60. Jiang, Y. *et al.* Electron ptychography of 2D materials to deep sub-ångström resolution. *Nature* **559**, 343–349 (2018).
61. Vincent, R. & Midgley, P. A. Double conical beam-rocking system for measurement of integrated electron diffraction intensities. *Ultramicroscopy* **53**, 271–282 (1994).
62. Midgley, P. A. & Eggeman, A. S. Precession electron diffraction - A topical review. *IUCrJ* **2**, 126–136 (2015).
63. Own, C., Dellby, N., Krivanek, O., Marks, L. & Murfitt, M. Aberration-corrected Precession Electron Diffraction. *Microsc. Microanal.* **13**, 96–97 (2007).
64. Savitzky, B. H. *et al.* Py4DSTEM: A Software Package for Four-Dimensional Scanning Transmission Electron Microscopy Data Analysis. *Microsc. Microanal.* **27**, 712–743 (2021).
65. Levin, B. D. A. Direct detectors and their applications in electron microscopy for materials science. *JPhys Mater.* **4**, (2021).
66. Caswell, T. A. *et al.* A high-speed area detector for novel imaging techniques in a scanning transmission electron microscope. *Ultramicroscopy* **109**, 304–311 (2009).
67. Fam, Y. *et al.* A versatile nanoreactor for complementary in situ X-ray and electron microscopy studies in catalysis and materials science. *J. Synchrotron Radiat.* **26**, 1769–1781 (2019).
68. Ma, C., Chen, L., Cao, C. & Li, X. Nanoparticle-induced unusual melting and solidification behaviours of metals. *Nat. Commun.* **8**, 1–7 (2017).
69. Huang, Y. *et al.* Effect of Ni on electrical properties of Ba(Zr,Ce,Y)O3-δ as electrolyte for protonic ceramic fuel cells. *Solid State Ionics* **390**, 116113 (2023).
70. Zhang, F. *et al.* Data synchronization in operando gas and heating TEM. *Ultramicroscopy* **238**, 113549 (2022).
71. Rauch, E. F. *et al.* Automated nanocrystal orientation and phase mapping in the transmission electron microscope on the basis of precession electron diffraction. *Zeitschrift fur Krist.* **225**, 103–109 (2010).
72. Ma, Z. D. G. S. *Metrology and Diagnostic Techniques for Nanoelectronics*. (2009).
73. Rouvière, J. L. & Sarigiannidou, E. Theoretical discussions on the geometrical phase analysis. *Ultramicroscopy* **106**, 1–17 (2005).
74. Schneider, C. A., Rasband, W. S. & Eliceiri, K. W. NIH Image to ImageJ: 25 years of image analysis. *Nat. Methods* **9**, 671–675 (2012).
75. Viladot, D. *et al.* Orientation and phase mapping in the transmission electron microscope using precession-assisted diffraction spot recognition: State-of-the-art results. *J. Microsc.* **252**, 23–34 (2013).
76. Madsen, J. & Susi, T. The abTEM code: transmission electron microscopy from first principles. *Open Res. Eur.* **1**, 24 (2021).
77. Giannuzzi, L. A. & Stevie, F. A. A review of focused ion beam milling techniques for TEM specimen preparation. *Micron* **30**, 197–204 (1999).
78. Giannuzzi, L. A., Drown, J. L., Brown, S. R., Irwin, R. B. & Stevie, F. A. Applications of the FIB lift-out technique for TEM specimen preparation. *Microsc. Res. Tech.* **41**, 285–290 (1998).
79. Wirth, R. Focused Ion Beam (FIB) combined with SEM and TEM: Advanced analytical tools for studies of chemical composition, microstructure and crystal structure in geomaterials on a nanometre scale. *Chem. Geol.* **261**, 217–229 (2009).
80. Wang, H., Xiao, S., Xu, Q., Zhang, T. & Zandbergen, H. Fast preparation of ultrathin FIB lamellas for MEMs-based in situ TEM experiments. *Mater. Sci. Forum* **850**, 722–727 (2016).

81. Gasser, P., Klotz, U. E., Khalid, F. A. & Beffort, O. Site-specific specimen preparation by focused ion beam milling for transmission electron microscopy of metal matrix composites. *Microsc. Microanal.* **10**, 311–316 (2004).
82. Hu, X., Koo, K., Smeets, P. J. M. & Dravid, V. P. Effects of Membrane Thickness, Gas Pressure and Electron Dose in Gas Cell Transmission Electron Microscopy. *Microsc. Microanal.* **29**, 1606–1607 (2023).
83. Barnard, J. S., Eggeman, A. S., Sharp, J., White, T. A. & Midgley, P. A. Dislocation electron tomography and precession electron diffraction - Minimising the effects of dynamical interactions in real and reciprocal space. *Philos. Mag.* **90**, 4711–4730 (2010).
84. Taylor, C. A., Nenoff, T. M., Pratt, S. H. & Hattar, K. Synthesis of complex rare earth nanostructures using: In situ liquid cell transmission electron microscopy. *Nanoscale Adv.* **1**, 2229–2239 (2019).
85. MacLaren, I. *et al.* A Comparison of a Direct Electron Detector and a High-Speed Video Camera for a Scanning Precession Electron Diffraction Phase and Orientation Mapping. *Microsc. Microanal.* **26**, 1110–1116 (2020).
86. Rebled, J. M., Yedra, L., Estradé, S., Portillo, J. & Peiró, F. A new approach for 3D reconstruction from bright field TEM imaging: Beam precession assisted electron tomography. *Ultramicroscopy* **111**, 1504–1511 (2011).
87. Rauch, E. F. & Véron, M. Crystal Orientation Angular Resolution with Precession Electron Diffraction. *Microsc. Microanal.* **22**, 500–501 (2016).
88. Mahr, C. *et al.* Theoretical study of precision and accuracy of strain analysis by nano-beam electron diffraction. *Ultramicroscopy* **158**, 38–48 (2015).
89. Hanlon, S. M., Persaud, S. Y., Long, F., Korinek, A. & Daymond, M. R. A solution to FIB induced artefact hydrides in Zr alloys. *J. Nucl. Mater.* **515**, 122–134 (2019).
90. Qiao, Y., Li, F., Li, S., Wang, Y. & Zhang, Y. The characterization of FIB-induced ζ-hydride in pure zirconium by HRTEM. *Mater. Lett.* **320**, 1–4 (2022).
91. Konishi, H., Ishikawa, A., Jiang, Y. B., Buseck, P. & Xu, H. Sealed environmental cell microscopy. *Microsc. Microanal.* **9**, 902–903 (2003).
92. Kawasaki, T., Ueda, K., Ichihashi, M. & Tanji, T. Improvement of windowed type environmental-cell transmission electron microscope for in situ observation of gas-solid interactions. *Rev. Sci. Instrum.* **80**, (2009).
93. Blume, R. *et al.* Monitoring in situ catalytically active states of Ru catalysts for different methanol oxidation pathways. *Phys. Chem. Chem. Phys.* **9**, 3648–3657 (2007).
94. Alan, T. *et al.* Characterization of Ultrathin Membranes to Enable TEM Observation of Gas Reactions at High Pressures. in *Volume 11: Mechanics of Solids, Structures and Fluids* 327–331 (ASMEDC, 2009). doi:10.1115/IMECE2009-11773.
95. Benner, G., Niebel, H. & Pavia, G. Nano beam diffraction and precession in an energy filtered CS corrected transmission electron microscope. *Cryst. Res. Technol.* **46**, 580–588 (2011).
96. Wang, L. One-Dimensional Electrical Contact to a Two-Dimensional Material. *Sci. (New York, N.Y.)432* **342**, 614–617 (2013).
97. Smith, J., Huang, Z., Gao, W., Zhang, G. & Chi, M. Atomic Resolution Cryogenic 4D-STEM Imaging via Robust Distortion Correction. *ACS Nano* **17**, 11327–11334 (2023).
98. Vermaak, N., Parry, G., Estevez, R. & Bréchet, Y. New insight into crack formation during corrosion of zirconium-based metal-oxide systems. *Acta Mater.* **61**, 4374–4383 (2013).
99. Fraker, A. C. *Corrosion of zircaloy spent fuel cladding in a repository*. (1989) doi:10.6028/NIST.IR.89-4114.
100. Ahmed, T. & Keys, L. H. The breakaway oxidation of zirconium and its alloys a review. *J. Less-Common Met.* **39**, 99–107 (1975).
101. Annand, K., Nord, M., MacLaren, I. & Gass, M. The corrosion of Zr(Fe, Cr)2 and Zr2Fe secondary phase particles in Zircaloy-4 under 350 °C pressurised water conditions. *Corros. Sci.* **128**, 213–223 (2017).

102. Ni, N. *et al.* Porosity in oxides on zirconium fuel cladding alloys, and its importance in controlling oxidation rates. *Scr. Mater.* **62**, 564–567 (2010).
103. Zhou, G. & Yang, J. C. Initial oxidation kinetics of Cu(100), (110), and (111) thin films investigated by in situ ultra-high-vacuum transmission electron microscopy. *J. Mater. Res.* **20**, 1684–1694 (2005).
104. Hu, J. *et al.* Identifying suboxide grains at the metal-oxide interface of a corroded Zr-1.0%Nb alloy using (S)TEM, transmission-EBSD and EELS. *Micron* **69**, 35–42 (2015).
105. Trimby, P. W. Orientation mapping of nanostructured materials using transmission Kikuchi diffraction in the scanning electron microscope. *Ultramicroscopy* **120**, 16–24 (2012).
106. Garner, A. *et al.* The microstructure and microtexture of zirconium oxide films studied by transmission electron backscatter diffraction and automated crystal orientation mapping with transmission electron microscopy. *Acta Mater.* **80**, 159–171 (2014).
107. Zaefferer, S. A critical review of orientation microscopy in SEM and TEM. *Cryst. Res. Technol.* **46**, 607–628 (2011).
108. Yankova, M. S. *et al.* Untangling competition between epitaxial strain and growth stress through examination of variations in local oxidation. *Nat. Commun.* **14**, 1–15 (2023).
109. Liu, J. *et al.* In-situ TEM study of irradiation-induced damage mechanisms in monoclinic-ZrO2. *Acta Mater.* **199**, 429–442 (2020).
110. Kim, H. G., Kim, I. H., Park, J. Y., Yoo, S. J. & Kim, J. G. In situ heating TEM analysis of oxide layer formed on Zr-1.5Nb alloy. *J. Nucl. Mater.* **451**, 189–197 (2014).
111. Ritchie, I. G. & Atrens, A. The diffusion of oxygen in alpha-zirconium. *J. Nucl. Mater.* **67**, 254–264 (1977).
112. Holmberg, B., Dagerhamn, T. & Leppänen, K. X-Ray Studies on Solid Solutions of Oxygen in a-Zirconium. *Acta Chem. Scand.* **15**, 919–925 (1961).
113. Grovenor, C. R. M. *et al.* Mechanisms of Oxidation of Fuel Cladding Alloys Revealed by High Resolution APT, TEM and SIMS Analysis. *MRS Proc.* **1383**, mrsf11-1383-a07-05 (2012).
114. Nicholls, R. J. *et al.* Crystal structure of the ZrO phase at zirconium/zirconium oxide interfaces. *Adv. Eng. Mater.* **17**, 211–215 (2015).
115. Cao, P. *et al.* Carbon nanotube (CNT) metal composites exhibit greatly reduced radiation damage. *Acta Mater.* **203**, (2021).
116. Neelisetty, K. K. *et al.* Novel thin film lift-off process for in situ TEM tensile characterization. *Microsc. Microanal.* **27**, 216–217 (2021).
117. Giannuzzi, L. A. Enhancing Ex-Situ Lift-Out with EXpressLO. *Microsc. Microanal.* **19**, 906–907 (2013).
118. Gorji, S. *et al.* Nanowire facilitated transfer of sensitive TEM samples in a FIB. *Ultramicroscopy* **219**, 113075 (2020).
119. Vijayan, S., Aindow, M., Jinschek, J. R., Kujawa, S. & Greiser, J. TEM Specimen Preparation for In Situ Heating Experiments Using FIB . *Microsc. Microanal.* **23**, 294–295 (2017).
120. Kwon, Y., An, B. S., Shin, Y. J. & Yang, C. W. Method of Ga removal from a specimen on a microelectromechanical system-based chip for in-situ transmission electron microscopy. *Appl. Microsc.* **50**, 0–5 (2020).
121. Giannuzzi, L. A. *et al.* Cryo-EXLO Manipulation of FIB Specimens for Cryo-TEM. *Microsc. Microanal.* **29**, 145–154 (2023).
122. Xin, H. L., Niu, K., Alsem, D. H. & Zheng, H. In situ TEM study of catalytic nanoparticle reactions in atmospheric pressure gas environment. *Microsc. Microanal.* **19**, 1558–1568 (2013).
123. Cautaerts, N. *et al.* Free, flexible and fast: Orientation mapping using the multi-core and GPU-accelerated template matching capabilities in the Python-based open source 4D-STEM analysis toolbox Pyxem. *Ultramicroscopy* **237**, 113517 (2022).
124. Yuan, R., Zhang, J., He, L. & Zuo, J.-M. Machine Learning Based Precision Orientation and Strain Mapping from 4D Diffraction Datasets. *Microsc. Microanal.* **27**, 1276–1278 (2021).
125. Ni, H.-C., Yuan, R., Zhang, J. & Zuo, J.-M. Emerging Machine Learning-Based Data Analysis

Techniques and Algorithms for Exploiting 4D-STEM Datasets. *Microsc. Microanal.* **30**, 4–6 (2024).
126. Shi, C. *et al.* Uncovering material deformations via machine learning combined with four-dimensional scanning transmission electron microscopy. *npj Comput. Mater.* **8**, (2022).
127. Ophus, C. *et al.* 4D-STEM Analysis with the Open Source py4DSTEM and crystal4D Toolkits. *Microsc. Microanal.* **28**, 3054–3055 (2022).
128. Munshi, J. *et al.* Disentangling multiple scattering with deep learning: application to strain mapping from electron diffraction patterns. *npj Comput. Mater.* **8**, 1–15 (2022).
129. Zeltmann, S. E., Bhusal, H. P., Yan, A. & Ophus, C. Robust Strain Analysis of Complex Heterostructures by Whole Pattern Fitting. *Microsc. Microanal.* **30**, (2024).

Supporting information for

*In-Situ* **Nanometer-Resolution Strain and Orientation Mapping for Gas-Solid Reactions via Precession-Assisted Four-dimensional Scanning Transmission Electron Microscopy**

Yongwen Sun[1,†], Ying Han[1,†], Dan Zhou[2,3,*], Athanassios S. Galanis[4], Alejandro Gomez-Perez[4], Ke Wang[5], Stavros Nicolopoulos[4], Hugo Perez Garza[3], Yang Yang[1,5,*]

[1] Department of Engineering Science and Mechanics, The Pennsylvania State University, University Park, PA, 16802, United States.
[2] Leibniz-Institut für Kristallzüchtung (IKZ), Max-Born-Str. 2, 12489 Berlin, Germany.
[3] DENSsolutions B.V., Delft, South Holland, 2628 ZD, Netherlands.
[4] NanoMEGAS SPRL, Rue Émile Claus 49 bte 9, 1050 Brussels, Belgium.
[5] Materials Research Institute, The Pennsylvania State University, University Park, PA, 16802, United States.
[†] These authors contributed equally to this work.
[*] Email of corresponding authors:
  dan.zhou@ikz-berlin.de (D.Z.), yangyang@alum.mit.edu (Y.Y.)

**Effects of Gaussian blurring filter on strain maps**

The strain mapping results, as depicted in **Fig. S1 a-e**, exhibit notable differences between the PED-off and PED-on modes. Specifically, the PED-on mode yields smoother strain maps with reduced false positive or negative signals. This observation prompts a critical question: Is the observed improvement in accuracy due to a mere blurring effect (due to a worse spatial resolution under PED-on mode), or does the PED mode actually enhance measurement precision? To explore this, we compared strain mapping under three conditions: (i) PED-off mode without any filter, (ii) PED-off mode with Gaussian blurring applied at various σ values, and (iii) PED-on mode without a filter. The black dotted lines mark regions where false strain signals are observed in the PED-off condition (**Fig. S1a**). The improvements in image quality introduced by Gaussian blurring and PED differ notably in the highlighted regions. In many highlighted regions, the sign of the measured strain is inverted after applying PED (e.g., switching from tensile to compressive), whereas Gaussian blurring alone rarely alters the strain sign. This comparison indicates that Gaussian blurring cannot effectively eliminate falsely detected strain signals, suggesting that the enhancement of strain accuracy by PED is physical.

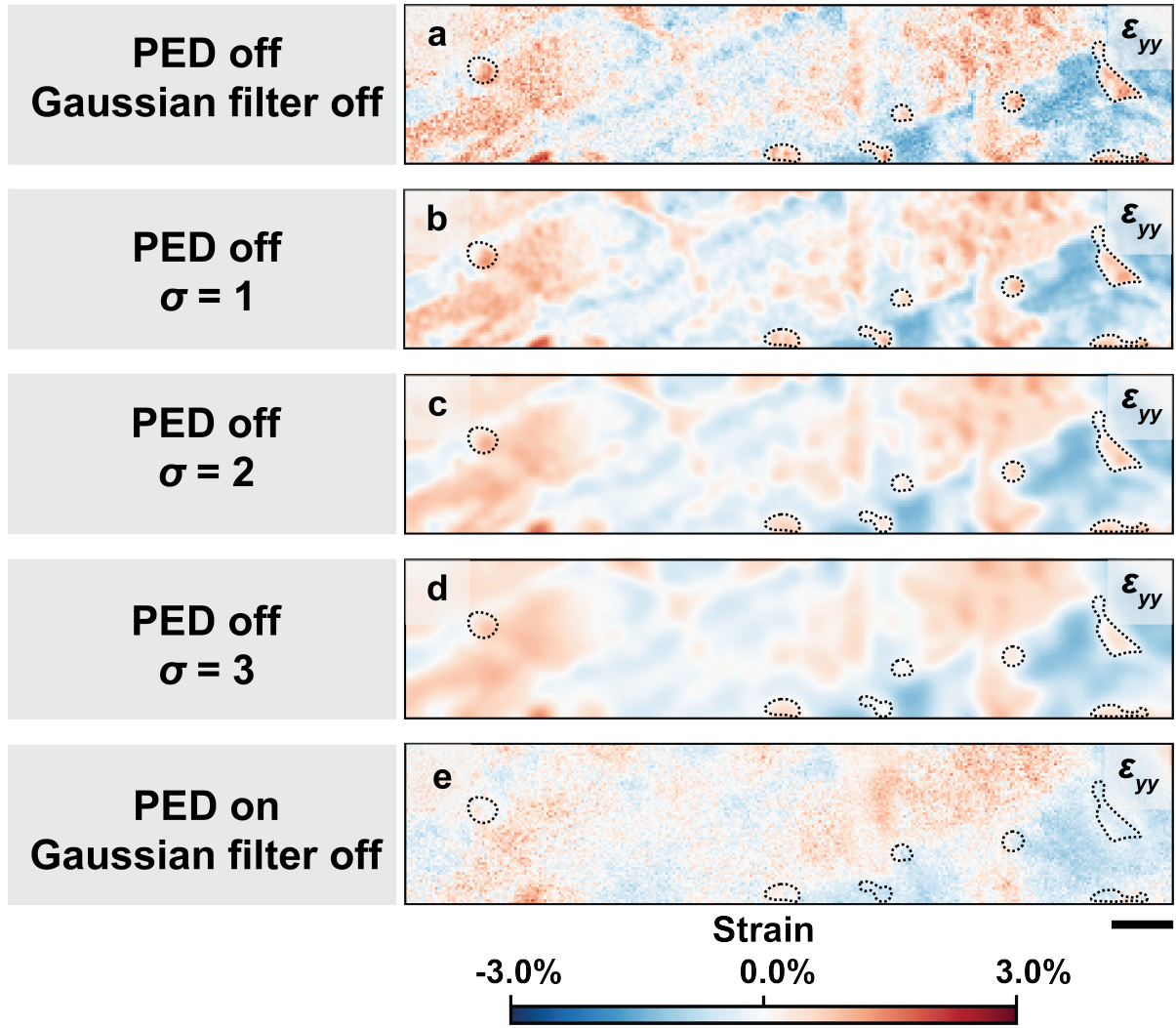

**Fig. S1.** Effects of Gaussian blurring filter on PED-off mode strain mapping in comparison with the pure PED-off and PED-on mode. **Only** $\varepsilon_{yy}$ strain maps are shown. Scale bars: 300 nm.

**Visibility enhancement of orientation mapping**

To highlight the variation of orientation in our sample, we have used a cropped version of inverse pole figures, as shown in **Fig. S2**.

The angle in hexagonal structure system between two crystal planes ($h_1$, $k_1$, $l_1$) and ($h_2$, $k_2$, $l_2$) can be calculated using the equation[1] below, and the *c*/*a* ratio of zirconium is taken as 1.593:

$$\cos\theta = \frac{h_1 h_2 + k_1 k_2 + 0.5(h_1 k_2 + k_1 h_2) + \frac{3a^2 l_1 l_2}{4c^2}}{\sqrt{(h_1^2 + k_1^2 + k_1 h_1 + \frac{3a^2 l_1^2}{4c^2})(h_2^2 + k_2^2 + k_2 h_2 + \frac{3a^2 l_2^2}{4c^2})}}$$

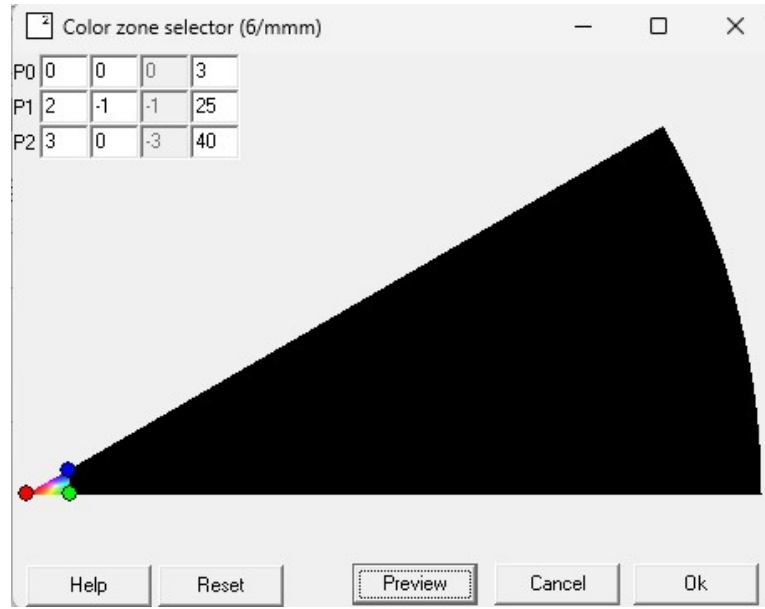

**Fig. S2**. Screenshot from ASTAR software illustrating the cropped inverse pole figure presented in **Fig. 2** of the manuscript.

**Reference:**

1. Hogan, D. W. & Dyson, D. J. Angles between planes in the hexagonal and tetragonal crystal systems. *Micron (1969)* **2**, 59–61 (1970).